\documentclass[a4paper,fleqn,usenatbib]{mnras}

\usepackage[T1]{fontenc}
\usepackage{ae,aecompl}

\usepackage{graphicx}	
\usepackage{amsmath}	
\usepackage{amssymb}	
\usepackage{comment}
\usepackage{lineno}
\usepackage{color}
\usepackage[toc,page]{appendix}

\newcommand{\ttwo}{{\tt Tempo2}}
\newcommand{\tn}{{\tt TempoNest}}

\title[Noise properties of 42 EPTA MSPs]{The noise properties of 42 millisecond pulsars from the European Pulsar Timing Array and their impact on gravitational wave searches}

\author[R. N. Caballero et al.]{\parbox{\textwidth}{R.~N.~Caballero,$^{1}$\thanks{E-mail:caball@mpifr-bonn.mpg.de}
K.~J.~Lee$^{2,1}$,
L.~Lentati$^{3}$,
G.~Desvignes$^{1}$,  
D.~J.~Champion$^{1}$,
J.~P.~W.~Verbiest$^{4,1}$,
G.~H.~Janssen$^{5,6}$,
B.~W.~Stappers$^{6}$,
M.~Kramer$^{1,6}$,
P.~Lazarus$^{1}$,
A.~Possenti$^{7}$,
C.~Tiburzi$^{4,1}$,
D.~Perrodin$^{7}$,
S.~Os{\l}owski$^{4,1}$,
S.~Babak$^{8}$,
C. G. Bassa$^{5}$,
P.~Brem$^{8}$,
M.~Burgay$^{7}$,
I.~Cognard$^{9,10}$,
J.~R.~Gair$^{11}$,
E.~Graikou$^{1}$,
L.~Guillemot$^{9,10}$,
J.~W.~T.~Hessels$^{5,12}$,
R.~Karuppusamy$^{1}$,
A.~Lassus$^{1}$,
K.~Liu$^{1}$,
J.~McKee$^{6}$,
C.~M.~F.~Mingarelli$^{13,1}$,
A.~Petiteau$^{14}$,
M.~B.~Purver$^{6}$,
P.~A.~Rosado$^{15,16}$,
S.~Sanidas$^{12,6}$,
A.~Sesana$^{17, 8}$,
G. Shaifullah$^{4,1}$,
R.~Smits$^{5}$,
S.~R.~Taylor$^{18}$,
G. Theureau$^{9,10}$,
R.~van~Haasteren$^{18}$
and A. Vecchio$^{17}$
  }
\vspace{0.4cm} \\ 
\parbox{\textwidth}{ 
$^{1}$Max-Planck-Institut f{\"u}r Radioastronomie, Auf dem H{\"u}gel 69, 53121 Bonn, Germany\\
$^{2}$Kavli institute for Astronomy and Astrophysics, Peking University, Beijing 100871,P.R.China\\
$^{3}$Astrophysics Group, Cavendish Laboratory, JJ Thomson Avenue, Cambridge, CB3 0HE, UK\\
$^{4}$Fakult\"at f\"ur Physik, Universit\"at Bielefeld, Postfach 100131, 33501 Bielefeld, Germany\\
$^{5}$ASTRON, the Netherlands Institute for Radio Astronomy, Postbus 2, 7990 AA, Dwingeloo, The Netherlands \\
$^{6}$Jodrell Bank Centre for Astrophysics, School of Physics and Astronomy, The University of Manchester, Manchester M13 9PL,UK\\
$^{7}$INAF - Osservatorio Astronomico di Cagliari, via della Scienza 5, I-09047 Selargius (CA), Italy\\
$^{8}$Max-Planck-Institut f{\"u}r Gravitationsphysik, Albert Einstein Institut, Am M{\"u}hlenberg 1, 14476 Golm, Germany \\
$^{9}$ Laboratoire de Physique et Chimie de l'Environnement et de l'Espace LPC2E CNRS-Universit{\'e} d'Orl{\'e}ans, F-45071, Orl{\'e}ans, France\\
$^{10}$ Station de radioastronomie de Nan{\c c}ay, Observatoire de Paris, CNRS/INSU F-18330 Nan{\c c}ay, France\\
$^{11}$ School of Mathematics, University of Edinburgh, King's Buildings, Edinburgh EH9 3JZ, UK \\
$^{12}$ Anton Pannekoek Institute for Astronomy, University of Amsterdam, Science Park 904, 1098 XH Amsterdam, The Netherlands\\
$^{13}$ TAPIR, California Institute of Technology MC 350-17, Pasadena, California 91125, USA\\
$^{14}$ Universit\'e Paris-Diderot-Paris7 APC - UFR de Physique, Batiment Condorcet, 10 rue Alice Domont et L\'eonie Duquet 75205 PARIS CEDEX 13, France \\
$^{15}$  Centre for Astrophysics \& Supercomputing, Swinburne University of Technology, PO Box 218, Hawthorn VIC 3122, Australia\\
$^{16}$ Max-Planck-Institut f{\"u}r Gravitationsphysik, Albert-Einstein-Institut, Callinstra\ss e 38, 30167, Hanover, Germany \\
$^{17}$ School of Physics and Astronomy, The University of Birmingham, Edgbaston, Birmingham, B15 2TT, UK \\
$^{18}$Jet Propulsion Laboratory, California Institute of Technology, 4800 Oak Grove Drive, Pasadena, CA 91106, USA\\
$^{19}$Laboratoire Univers et Th\'eories LUTh, Observatoire de Paris, CNRS/INSU, Universit{\'e} Paris Diderot, 5 place Jules Janssen, 92190 Meudon, France\\
}}

\pubyear{2016}

\begin{document}
\label{firstpage}
\pagerange{\pageref{firstpage}--\pageref{lastpage}}
\maketitle

\begin{abstract}
The sensitivity of Pulsar Timing Arrays 
to gravitational waves depends 
on the noise present in the individual 
pulsar timing data. Noise may be either intrinsic or extrinsic to the pulsar. 
Intrinsic sources of noise 
will include rotational instabilities, for example. Extrinsic sources of noise include contributions from
physical processes which are not sufficiently well modelled, for example, dispersion and scattering effects,  
analysis errors and instrumental instabilities.
We present the results from a noise analysis for 42 millisecond pulsars (MSPs) 
observed with the European Pulsar Timing Array. 
For characterising the low-frequency, stochastic and 
achromatic noise component, 
or ``timing noise'', 
we employ two methods, based on Bayesian 
and frequentist statistics. 
For 25 MSPs, we achieve statistically significant measurements of 
their timing noise parameters and 
find that the two methods give consistent results. For the remaining 17 MSPs, we place 
upper limits on the timing noise amplitude at the 95\% confidence level.
We additionally place an upper limit on the contribution to the pulsar noise budget 
from errors in the reference terrestrial time standards (below 1\%), 
and we find evidence for a noise component which is present only 
in the data of one of the 
four used telescopes. 
Finally, we estimate that the timing noise of individual 
pulsars reduces the sensitivity of this data set to an 
isotropic, stochastic GW background by a factor of $>9.1$ 
and by a factor of $>2.3$ for 
continuous GWs from resolvable, 
inspiralling supermassive black-hole binaries with circular orbits.
\end{abstract}




\section{Introduction}
\label{sect:intro}
Over the past decades, pulsar astronomy has been instrumental 
in the experimental tests of general relativity (GR) and alternative theories of gravity.
Some of the most notable highlights from this research field include 
the first evidence of the existence of gravitational waves 
(GWs) \citep{1989ApJ...345..434T}, the most precise  
tests of GR \citep{2006Sci...314...97K}, as well as tests of
alternative theories of gravity, such as tensor-scalar gravity, 
in the quasi-stationary, strong-field regime 
\citep[see e.g.][]{2012MNRAS.423.3328F,
2013CQGra..30p5019S}.
These results rely on the pulsar timing technique 
\citep[e.g.][]{2005hpa..book.....L}, which fits the precisely recorded times-of-arrival (TOAs) of 
the pulses with a model of the pulsar's rotational, astrometric and orbital parameters, 
as well as signal propagation delays induced by the ionised
interstellar medium between the pulsar and Earth. The 
differences between the observed TOAs and those predicted by the model are 
called the timing residuals and contain the effects of 
any unmodelled physical or instrumental processes.

One of the applications of pulsar timing is the possibility of 
direct detection of GWs via the precise timing of 
an ensemble of pulsars, commonly referred to as 
a Pulsar Timing Array \citep[PTA; ][]{1990ApJ...361..300F}. 
The expected effects of GW propagation on the TOAs were first examined by \cite{1978SvA....22...36S}. 
Later, the idea of using a PTA for unambiguous direct detection of low-frequency (nHz regime) GWs
based on the predicted cross-correlation of the residuals of pulsars
in various sky positions was proposed by \cite{1983ApJ...265L..39H}. 
Subsequent work has identified the 
potential of modern timing data for detecting nHz GWs and 
formulated the detection methodologies \citep[e.g.][]{2004ApJ...606..799J,2005ApJ...625L.123J,2012PhRvD..85l2003S}.

PTAs are sensitive to the stochastic GW background (GWB) 
resulting from the incoherent superposition of the 
GW signals from the cosmic population 
of unresolved inspiralling 
supermassive black-hole binaries (SMBHBs) 
\citep[e.g.][]{1995ApJ...446..543R}, continuous GWs (CGWs) from individual, resolvable  
SMBHB systems \citep[e.g.][]{1975GReGr...6..439E}, the GWB created from the 
decaying loops of a cosmic string network that may have formed in the early Universe
\citep[e.g.][]{1976JPhA....9.1387K}, 
a cosmological relic GWB from the Universe's inflationary era 
\citep[e.g.][]{2005PhyU...48.1235G} and 
the memory term (long-term change in the GW's amplitude) 
from GW bursts from SMBHB mergers 
\citep[e.g.][]{2009PhRvD..80b4002F}. Prior to the detection, upper limits on the GW amplitudes can 
impose limits on the properties of the cosmic SMBHB population \citep[e.g][]{2015Sci...349.1522S},
and rule out the presence of nearby SMBHBs proposed by independent 
observations \citep{2004ApJ...606..799J}. In the era of GW astronomy, PTAs using 
future, hyper-sensitive telescopes will also
be able to test theories of gravity in the radiative regime. The 
GW polarisation modes predicted by GR or alternative theories result 
in different spatial cross-correlations of the pulsar timing residuals 
\citep[e.g.][]{2012PhRvD..85h2001C}. These cross-correlations 
can be further modified if the graviton is not massless as predicted by GR \citep[e.g.][]{2013CQGra..30v4016L}.

The pursuit of GW detection using pulsar timing is co-ordinated by three 
consortia; the European Pulsar Timing Array \citep[EPTA; ][]{2013CQGra..30v4009K} in Europe, 
the North-American Nanohertz Observatory for 
Gravitational Waves \citep[NANOGrav; ][]{2013CQGra..30v4008M} 
in North America and the Parkes Pulsar Timing Array \citep[PPTA; ][]{2013CQGra..30v4007H} in Australia. 
The PTAs employ in total eight large single-dish radio telescopes. The EPTA uses five telescopes, namely the 
Effelsberg Radio Telescope, the Nan\c cay Radio Telescope, the Lovell Telescope, the 
Westerbork Radio Synthesis Telescope and the Sardinia Radio Telescope. NANOGrav uses two 
telescopes, the Green Bank Telescope and the Arecibo Radio Telescope, while the 
PPTA uses the Parkes Radio Telescope.
The three consortia co-operate under the International Pulsar Timing Array 
(IPTA) consortium, maximising the observing efficiency and data set sensitivity. 

The sensitivity of a given PTA is mainly 
limited by the uncertainties of the TOA measurements, 
the number of observations and the data time-span,
the number of pulsars, their sky distribution and the presence of 
low-frequency noise in the data \citep[see e.g.][]{2012MNRAS.423.2642L,2013CQGra..30v4015S}. 
While improvements in the 
instrumentation, increase of the allocated telescope time to PTAs and 
discoveries of new pulsars can address the first three factors, low-frequency 
noise needs to be characterised and understood on a pulsar-by-pulsar basis. 

A number of methods have been developed to 
mitigate the dominant sources of noise in pulsar timing. 
Temporal variations in the dispersion measure 
(DM; integrated free electron density of the interstellar medium) 
along the line of sight to the pulsar is a primary source 
of low-frequency stochastic noise.
DM time delays, however, depend 
on the observing frequency, $\nu$, as $t_{DM}\propto DM \nu^{-2}$, and 
therefore DM variations can be, to a large degree, 
corrected using multi-frequency data, 
\citep[e.g.][]{2013MNRAS.429.2161K,2014MNRAS.441.2831L}. 
Improper calibration of the gain of the two receiver feeds or cross-coupling 
between the two feeds can potentially lead to distortions of the total intensity profiles.
These instrumental artefacts will introduce additional non-stationary noise components in the timing residuals
 \citep{2003ASPC..302...65V,2006ApJ...642.1004V}. By performing standard calibration 
 observations during every observing run, we can minimise the presence of such noise in the data 
 \citep[e.g.][]{2000ApJ...532.1240B}. By comparing the noise properties of the same pulsars 
using overlapping data from from different telescopes, 
uncorrected noise from instrumental instabilities can potentially be identified (Lentati et al., submitted).

Unfortunately, pulsar timing data also exhibit 
some levels of ``timing noise'' (TN), 
low-frequency, stochastic, achromatic noise, 
the physical origin of which is unknown and, 
as such, cannot be mitigated. 
TN is primarily thought to be caused by pulsar rotational instabilities from  
various mechanisms. 
One approach is to consider simultaneous random walks and discrete jumps 
(caused, e.g., by micro-glitches) in the pulsar's spin frequency and the spin-down rate 
\citep[e.g.][]{1985ApJS...59..343C,1995MNRAS.277.1033D,2010ApJ...725.1607S}. 
Based on observational evidence, it is also 
suggested that TN can result from accumulated periodic and quasi-periodic 
changes in the spin-down rates due to magnetospheric state switching 
\citep{2006Sci...312..549K,2010Sci...329..408L}.  
In addition, intrinsic noise has also been proposed to be the result of 
undetected (and therefore unmodelled) bodies in orbit, 
such as asteroid belts \citep{2013ApJ...766....5S} or planetary-mass objects 
in long, decadal orbits \citep{1999ApJ...523..763T}. Clearly, the measured TN in pulsar timing 
data can be a superposition of noise intrinsic to the pulsar, and any of the above 
non-intrinsic noise which is not properly mitigated, 
e.g. noise by DM variations not properly corrected 
due to the lack of sufficient multi-frequency data.

While young pulsars show 
large amounts of low-frequency noise, millisecond pulsars (MSPs), 
typically show very low levels of such noise \citep{2009MNRAS.400..951V}. 
It is theorised that MSPs have spun-up to 
the observed ms-order rotational periods via mass transfer from their 
companions during the system's evolution \citep{1982Natur.300..728A}. 
Their highly stable rotations, short periods and absence of significant temporal 
changes in their pulse profile shapes \citep[see e.g.][]{2013CQGra..30p5019S} 
make them excellent celestial clocks 
which can be timed to sub-100~ns precision over decades. 
MSPs are therefore the 
observed sources for GW-detection experiments, 
and indeed for all high-precision pulsar timing applications. 

Despite their demonstrated rotational stability, some MSPs 
show significant amounts of TN. While their TN is considerably weaker than 
that of non-recycled pulsars, it can be significant enough to hinder GW detection.
PSR B1937+21 (J1939+2134), the first ever discovered MSP, is a notable example 
of an MSP with strong TN \citep{1994ApJ...428..713K,2013ApJ...766....5S}. 
Other MSPs show more moderate noise levels, comparable to 
the predicted strength of the targeted GWs 
signals \citep[e.g. PSR J1713+0747; see][]{2015ApJ...809...41Z}. 
The characterisation of TN is therefore of central importance 
in high-precision pulsar timing applications. 

The measured TN will also contain signals from spatially correlated 
low-frequency noise \citep[e.g.][]{2016MNRAS.455.4339T}. Primary examples 
are the long sought-after stochastic GWB, the signal caused by errors 
in the reference terrestrial time standards \citep[see e.g.][]{2012MNRAS.427.2780H} 
and errors in the solar system ephemeris \citep[see][]{2010ApJ...720L.201C}. These 
signals can be distinguished by the spatial cross-correlations they induce on the 
timing residuals. The GWB induces a quadrupole signature 
(see Section~\ref{subsec:GWB_limits}). 
Errors in the terrestrial time standards produce a fully correlated signal in all pulsars (see Section~\ref{sect:clockErr})
while errors in the solar system ephemeris can potentially produce a superposition of
dipolar correlations between pulsars, each produced by the error in the predicted location of a solar system body. 
PTAs allow such correlated signals to be recovered or put upper limits on their power.

Different methods have been proposed and employed to characterise the statistical 
properties of TN in pulsar data and to perform pulsar timing analysis in the presence of correlated noise. 
These cover techniques based on frequentist \citep{1997A&A...326..924M,2011MNRAS.418..561C} 
and Bayesian statistics \citep[e.g.][]{2009MNRAS.395.1005V, 2014MNRAS.437.3004L}, 
both in the time- and frequency-domain. 
As part of the efforts to detect GWs, an increasing number of algorithms
are being used by the various PTAs to determine the TN properties of
MSPs, motivating work to examine the possible biases inherent
to different methods. 
In this context, we perform characterisation of the TN using two established methods 
based on different statistical analyses, Bayesian and frequentist, and 
make a comparison of their performance and results. 
We subsequently use the measured TN properties to 
search for the presence of TN unique to specific observing systems, 
place an upper limit on the contribution of clock errors to the measured noise 
and investigate the impact of the TN on the data set's sensitivity to GWs. 

This paper is organised as follows. In Section~\ref{sect:Dataset}, we describe the data we use. 
In Section~\ref{sect:Methods}, we present the methods used to calculate 
the noise parameters. The results from both methods are presented in Section~\ref{sect:results}. 
In Section~\ref{sect:systematics}, we check for TN present only in individual data subsets and continue to investigate systematics by making a search for a correlated clock error signal in Section~\ref{sect:clockErr}.
In Section~\ref{sect:GWprospect}, we evaluate the effects of the TN present in our data on their 
sensitivity to GWBs and CGWs and finally discuss our conclusions in Section~\ref{sect:Conclusions}.

\section[]{The EPTA Legacy Timing Dataset}
\label{sect:Dataset}
We use the EPTA Legacy data set that is presented in 
Desvignes et al., (submitted; henceforth D15). The data set is composed of 
data recorded with four EPTA radio telescopes: The Effelsberg Radio Telescope (EFF) in Germany, 
the Nan\c cay Radio Telescope (NRT) in France, 
the Westerbork Synthesis Radio Telescope (WSRT) in the Netherlands 
and the Lovell Telescope (JBO) in the United Kingdom. 
The data-recording systems (backends) used are the Effelsberg-
Berkeley Pulsar Processor (EBPP), the Berkeley-Orl\'eans-Nan\c cay (BON), 
the Pulsar Machine I (PuMaI) and the Digital 
Filterbank (DFB) respectively. A more 
detailed description of the instruments and data reduction techniques can 
be found in D15, where the timing solutions of the pulsars 
are also presented. 

The data set includes TOAs from 42 MSPs. Their key properties are 
summarised in Table~\ref{tab:character}.
We identify observing systems as
unique combinations of telescope, backend and 
central observing frequency (receiver). In total, the data set has 18 
distinct systems. The EBPP L-band\footnote{1 to 2 GHz range in centre frequency} data have 
the longest time-span, with a maximum of 18 years, starting from October 1996, 
divided into two observing systems, due to a change in the receiver in 2009. 
For most of the sources with EBPP data, all other instruments started recording from 2007 onwards, 
dividing our longest pulsar data sets into two subsets: the first, with single-telescope, single-frequency 
data and the second, with multi-telescope, multi-frequency data. 
The lack of multi-frequency data in the first half of the 
data set makes direct measurements and corrections of the DM variations impossible. 
It is however possible to extrapolate the signal measured in 
the second epoch to the first, under the assumption that 
the DM variations signal is stationary \citep[see][]{2014MNRAS.441.2831L}. This is performed using 
the Bayesian analysis methods described in Section~\ref{sect:Bayes}. 
For a number of MSPs (e.g. PSR J1713+0747, PSR J1012+5307), 
multi-telescope coverage begins in 1999 with 
PuMaI data, which contain good quality low-frequency data, 
allowing direct measurements of the DM variations almost throughout the data set.
We note that four MSPs (see Table~\ref{tab:character}) 
suffer from a gap in the Effelsberg L-band 
data for the period
between April 1999 and October 2005. 
The gap is due to changes in the observing priorities. 

\begin{table}
\centering
\caption{General characteristics of the EPTA Legacy data set. For each pulsar we note the total time-span, T, the ranges 
of the observing frequencies, $\nu$, the number of observing systems and the number of TOAs. 
Sources marked with a star suffer from a gap of $\sim$6 years (1999-2005) in the
Effelsberg 1410 MHz data.}
\begin{tabular}{l c c c c}
\hline\hline
PSR       & T   & $\nu$ range &  number of  & number of \\
J-Name       &   (yrs)           & (MHz) &  systems & TOAs \\
\hline
J0030+0451$^{\star}$ & 15.1 & 1345-2678 & 7 & 907\\
J0034$-$0534 & 13.5 & 323-1628 & 6 & 276\\
J0218+4232 & 17.6 & 323-2683 & 13 & 1196\\
J0610$-$2100 & 6.9 & 1365-1632 & 3 & 1034\\
J0613$-$0200 & 16.1 & 323-2636 & 14 & 1369\\
J0621+1002 & 11.8 & 323-2635 & 10 & 673\\
J0751+1807 & 17.6 & 1352-2695 & 9 & 796\\
J0900$-$3144 & 6.9 & 1365-2303 & 5 & 875\\
J1012+5307 & 16.8 & 323-2636 & 15 & 1459\\
J1022+1001 & 17.5 & 323-2634 & 10 & 908\\
J1024$-$0719$^{\star}$ & 17.3 & 1346-2628 & 9 & 561\\
J1455$-$3330 & 9.2 & 1367-1698 & 3 & 524\\
J1600$-$3053 & 7.6 & 1366-2298 & 4 & 531\\
J1640+2224 & 17.3 & 1335-2636 & 8 & 595\\
J1643$-$1224 & 17.3 & 1353-2639 & 11 & 759\\
J1713+0747 & 17.7 & 820-2637 & 14 & 1188\\
J1721$-$2457 & 12.7 & 1335-1698 & 4 & 150\\
J1730$-$2304$^{\star}$ & 16.7 & 1352-2629 & 8 & 268\\
J1738+0333 & 7.3 & 1366-1630 & 3 & 318\\
J1744$-$1134 & 17.3 & 323-2634 & 9 & 536\\
J1751$-$2857 & 8.3 & 1397-1631 & 3 & 144\\
J1801$-$1417 & 7.1 & 1395-1697 & 3 & 126\\
J1802$-$2124 & 7.2 & 1366-2048 & 4 & 522\\
J1804$-$2717 & 8.1 & 1374-1698 & 3 & 116\\
J1843$-$1113 & 10.1 & 1335-1629 & 5 & 224\\
J1853+1303 & 8.4 & 1397-1698 & 3 & 101\\
J1857+0943 & 17.3 & 1335-2632 & 9 & 444\\
J1909$-$3744 & 9.4 & 1367-2681 & 3 & 425\\
J1910+1256 & 8.5 & 1366-1630 & 3 & 112\\
J1911$-$1114 & 8.8 & 1397-1630 & 4 & 130\\
J1911+1347 & 7.5 & 1365-1698 & 3 & 140\\
J1918$-$0642 & 12.8 & 1372-1630 & 6 & 278\\
J1939+2134 & 24.1 & 820-2278 & 12 & 3172\\
J1955+2908 & 8.1 & 1395-1629 & 4 & 157\\
J2010$-$1323 & 7.4 & 1381-2298 & 5 & 390\\
J2019+2425 & 9.1 & 1365-1629 & 3 & 130\\
J2033+1734 & 7.9 & 1367-1631 & 4 & 194\\
J2124$-$3358 & 9.4 & 1365-2298 & 5 & 544\\
J2145$-$0750 & 17.5 & 323-2683 & 12 & 800\\
J2229+2643 & 8.2 & 1355-2637 & 6 & 316\\
J2317+1439$^{\star}$ & 17.3 & 1352-2637 & 8 & 555\\
J2322+2057 & 7.9 & 1395-1698 & 4 & 229\\
\hline
\hline
\end{tabular}
\label{tab:character}
\end{table}
 
\section[]{Methods for Estimating Noise Properties}
\label{sect:Methods}
 
For the estimation of the noise properties, we use two different methods. 
The first method follows a Bayesian approach, in the 
time-frequency domain and is described in \cite{2014MNRAS.437.3004L}. 
The second method uses frequentist statistics based on power-spectral 
estimation of the residuals and using algorithms described in Section~\ref{sect:Freq}, 
which are an extension of those introduced in \cite{2011MNRAS.418..561C}. 
We first discuss the 
noise model components, which we use for 
both approaches, and then present the details 
of each method used.

\subsection[]{Noise Modelling}
\label{sect:Budget} 
We form the timing residuals using the pulsar timing analysis 
package \ttwo{} \citep{2006MNRAS.369..655H}, which
iteratively performs a weighted least-squares (wLS) fit of the model to the TOAs 
until the reduced chi-squared of the residuals is minimised. 
Timing models are gradually improved over many years 
by incorporating more data. These solutions will often result in timing residuals scattered 
beyond what would be expected based on their formal uncertainties, due to the 
absence, at this point, of the stochastic signals in the model. These signals are in 
general divided into the time-correlated and uncorrelated components.

The uncorrelated (white-noise) components correct the uncertainties of the timing residuals.
The formal uncertainties of the TOAs are derived by the cross-correlation of the recorded 
integrated pulse profile with a reference template, which is constructed using the best available observations. 
These uncertainties are 
correct if the recorded profiles are characterised solely by (white) radiometer noise and the 
profile template precisely represents the intrinsic shape of the integrated profile. 
However, possible presence of un-excised 
radio frequency interference (RFI), temporal variations in the pulse profile, artefacts in the profiles 
from instrumental instabilities or imperfect 
profile templates can lead to errors in the uncertainty 
estimations \citep[e.g.][]{2011MNRAS.417.2916L}. 
It is therefore common practice 
to include a multiplicative correction factor called EFAC. We also add a correction term quadratically to the 
formal uncertainty to account for additional scatter in the TOAs caused by 
statistically independent physical processes, such as
pulse phase jitter noise \citep[e.g.][]{2014MNRAS.443.1463S}. This term is commonly referred to as EQUAD. 
We do not investigate the physical origin of the noise included in the EQUADs. 
This requires a more detailed analysis of the white noise; for example, 
jitter noise is dependent on the integration time of the
observation and this needs to be properly taken into consideration 
if one wants the EQUAD number to describe an 
underlying physical process.

We include one EFAC and one EQUAD term per observing system
to mathematically model the uncorrelated noise from all possible processes. 
The white-noise correction factors should be such that the data 
satisfy the central assumption of pulsar timing, that they are drawn from a random Gaussian process.
In other words, when subtracting the waveforms (induced residuals) of all 
calculated stochastic signals from 
the residuals, their uncertainties should be such that 
the residuals are white and the timing solution 
has a reduced chi-squared of unity.
The original TOA uncertainty, $\sigma$, EFAC (f), EQUAD (q) and corrected uncertainy, $ {\hat \sigma}$, are related\footnote{This definition is not unique. \ttwo{} by default defines the correction as ${\hat \sigma}^2 =f^2\cdot(\sigma^2+q^2)$} as:

\begin{center}
\begin{equation}
\label{eq:TRsigma}
{\hat \sigma}^2 = ( \sigma\cdot f)^2+q^2
\end{equation}
\end{center}   
We include two stationary time-correlated noise components, namely the chromatic low-frequency noise from DM variations
and the achromatic TN. Previous studies \citep[e.g.][]{2010ApJ...725.1607S,2011MNRAS.418..561C} 
have shown that 
the low-frequency power spectra of pulsar timing residuals 
can be adequately modelled with single power-laws for the majority of MSPs. 
This does not mean that the TN is necessarily a pure power-law, 
but rather that this functional form is sufficient to describe the data, given the measurement precision. 
We examined 
whether deviations from the single power-law model are supported by the data using the Bayesian analysis 
method. In particular, we performed the noise analysis with two additional models for the TN spectrum: 
(i) a model that allows the power of individual frequency bins to 
vary independently from the power law model and (ii) a model that includes the power-law and an additional 
sinusoid signal of varying frequency, amplitude and phase. 
We evaluated the results using the Bayes factor, i.e. the ratio of the 
Bayesian evidence of two competing models (see also Section~\ref{sect:Bayes}). 
A common interpretation of the Bayes factor is given by Kass \& Raftery (1995), based on which 
we required a value equal or greater than 3 to justify the addition of any extra model parameter.
This was not the case for any of the models we compared to the simple single power-law model.

In this work, we have followed the single power-law formalism 
for both analysis methods in order to facilitate their 
comparison and the comparison 
of the measured TN parameters with those usually used as GW stochastic parameters in the PTA literature.
For isotropic GW signals (see Section~\ref{sect:GWprospect}) 
one of the most important properties is the characteristic strain spectrum, $h_c(f)$, of the GWB on the 
one-sided power spectrum of the induced timing residuals. For most models of interest, 
this can be written as a power-law function of the GW frequency \citep[e.g.][]{2005ApJ...625L.123J}, $f$ as:
\begin{center}
\begin{equation}
\label{eq:GWBstrain}
h_c(f)=A\Big(\frac{f}{f_r}\Big)^\alpha
\end{equation}
\end{center}
where  $A$ is the (dimensionless) amplitude of the wave, $\alpha$ is the 
spectral index\footnote{We define the index positive, but note that in the literature it 
is sometimes defined as a negative number}
and $f_r$ is the reference frequency, typically set to 1~yr$^{-1}$. The one-sided 
power spectral density of the signal is then given by:
\begin{center}
\begin{equation}
\label{eq:RedNoiseFunc}
S(f)=\frac{A^2}{12\pi^2}\Big(\frac{f}{f_r} \Big)^{-\gamma}
\end{equation}
\end{center}
where the power spectrum and strain spectral indices are related as $\gamma\equiv 3-2\alpha$. 
This is the functional form we use to model the TN.
We set a cut-off at frequency 1/T, where T 
is the time-span of the data. The cut-off arises naturally due to the absorption of 
the lowest frequencies by the fitting of the pulsar's spin and spin-down in the timing model. 
It has been shown \citep{2012MNRAS.423.2642L,2013MNRAS.428.1147V} 
that if the spectral index is $\gamma\lesssim7$ 
(which is the case for all MSPs in this paper), 
the cut-off at frequency 1/T is sufficient.

The DM variations have been mitigated using first- and second-order DM derivatives in the timing 
model (which are first- and second-order polynomials) 
and additionally a power law equivalent to Eq.~\eqref{eq:RedNoiseFunc}. The DM derivatives absorb 
any DM variation signal with frequencies below the cut-off frequency, in the same way the spin 
and spin-down do for the achromatic TN \citep{2014MNRAS.441.2831L}. The observing 
frequency dependence of the DM variations signal is measured in the time-domain via 
the (multi-frequency) timing residuals, as we show in Section~\ref{sect:Bayes}.
The choice of a power-law spectrum for the DM variations 
is motivated by the fact that, across a wide spatial frequency 
range, the electron density fluctuation spectrum usually follows a power-law \citep{1995ApJ...443..209A}.

\subsection[]{Noise Parameter Estimation Using Bayesian Inference}
\label{sect:Bayes}

The first Bayesian investigation of the GWB detectability with PTAs was performed by
\cite{2009MNRAS.395.1005V}. The algorithms were later applied on EPTA data to derive 
the EPTA GWB upper limit \citep{2011MNRAS.414.3117V}.
In that analysis, the timing noise parameters of the MSPs were simultaneously estimated with the 
GWB parameters. Further work on Bayesian analysis methods for 
pulsar timing provided more algorithms, both in time- and time-frequency-domains, to characterise the properties 
of timing noise and DM variations and to perform robust pulsar timing analysis in the 
presence of correlated noise \citep[e.g.][]{2011MNRAS.414.3117V,2013PhRvD..87j4021L,2014MNRAS.441.2831L}. 

Bayes' theorem, which is the central equation for these analysis methods, states that:

 \begin{center}
 \begin{equation}
 \label{eq:Bayes}
 Pr(\Theta)=\frac{L(\Theta) \pi (\Theta)}{Z} ,
 \end{equation}
 \end{center}
where $\Theta$ is the model's parameters, 
$Pr(\Theta)$ is the posterior probability distribution (PPD)
of the parameters (probability distribution of the parameters given the model and the data), 
$\pi(\Theta)$ is the prior 
probability distribution (pPD) of the parameters 
(the initial hypothesis of the probability distribution of the parameters 
for a given model), $L(\Theta)$ is the likelihood function 
(which gives the probability that the data are described by a given model) 
and $Z$ is the Bayesian evidence. $Z$ is only
a normalising factor independent of $\Theta$ and can therefore be ignored when one is 
interested only in parameter 
estimation, such that $Pr(\Theta)\propto L(\Theta) \pi (\Theta)$. On the other hand, when 
one is interested in 
model selection, the ratio of the evidence between 
two different models, $\mathcal{R}$, known as the Bayes factor, is used. 
The probability, $\mathcal{P}$, of a model compared to another, 
can the expressed (Kass \& Raftery 1995) as:

\begin{center}
\begin{equation}
\label{eq:GWBstrain}
\mathcal{P}=\frac{\mathcal{R}}{1+\mathcal{R}}
\end{equation}
\end{center}
The various Bayesian analysis algorithms are distinguished by the mathematical description of the model 
parameters and the computational methods used to sample the unnormalised PPD.

\cite{2014MNRAS.437.3004L} introduced \tn{}, 
a Bayesian software package for the analysis of pulsar timing data, 
available to use as a \ttwo{} plug-in. 
The timing solution and the additional stochastic parameters such as EFACs, EQUADs, 
DM variations and the TN (referred to as ``excess red noise'') can be determined simultaneously. 
\tn{} uses the Bayesian inference tool MultiNest \citep{2008MNRAS.384..449F} 
to explore this joint parameter space, whilst using \ttwo{} as an established means of 
evaluating the timing model at each point in that space. For the PPD sampling, \tn{} uses 
the nested sampling Monte-Carlo method \citep{2004AIPC..735..395S}.

We perform a joint analysis for the timing model and the stochastic parameters. 
Both the TN and the DM variations 
are modelled as Gaussian stochastic signals with power-law spectra as described 
by Eq.~\eqref{eq:RedNoiseFunc}. 
\tn{} employs the
time-frequency analysis described in \cite{2013PhRvD..87j4021L}. The TN waveform is expressed as 
${\rm \bf t}_{\rm TN}=\mathbfss{F}_{\rm TN}\textrm{{\bf a}}$, where $\mathbfss{F}_{\rm TN}$ 
is the Fourier transform with components 
$\mathsf{F}=\textrm{sin}(2\pi f) + \textrm{cos}(2\pi f)$ and corresponding coefficients, 
$\textrm{{\bf a}}$, which are free parameters. 
Here, and henceforth, we use boldface characters in equations to denote matrices.
The Fourier frequencies take values $f=n/$T, with n integers ranging from 1 up to the value 
necessary to sample 
frequencies as high as 1/14 days$^{-1}$.
The covariance matrix of the TN is then described by the following equation 
\citep[see][]{2015MNRAS.453.2576L}:

\begin{eqnarray}
\label{eq:BayesCovRed}
{\bf C}_{\textrm{TN}}&=&{\bf C^{-1}_{\textrm{w}}} - {\bf 
C^{-1}_{\textrm{w}}}\mathbfss{F}_{\rm TN} \left[(\mathbfss{F}_{\rm TN})^{\textrm{{\bf T}}}{\bf 
C^{-1}_{\textrm{w}}}\mathbfss{F}_{\rm TN} + 
(\Psi^{\mathbf{\mathrm{SN}}})^{-1}\right]^{-1}  \nonumber 
\\
& &
(\mathbfss{F}_{\rm TN})^{\textrm{{\bf T}}}{\bf C^{-1}_{\textrm{w}}} .
\end{eqnarray}
Here, $\Psi=\langle \textrm{a}_i\textrm{a}_j\rangle$, is the covariance matrix of the Fourier coefficients 
and ${\bf C_{\textrm{w}}}$ is the covariance matrix 
of the white noise component, a diagonal matrix 
with the main diagonal populated by the residual uncertainties squared,
${\hat \sigma}^2$ (as in Eq.~\ref{eq:TRsigma}). 
The superscript ${\textrm{{\bf T}}}$ denotes the transpose of the matrix.

The covariance matrix for the DM variations, ${\bf C_{\textrm{DM}}}$, is equivalent to Eq.~\eqref{eq:BayesCovRed}, 
but including an observing frequency dependence. 
This is achieved by replacing $\mathbfss{F}$ components 
with ${\mathsf{F}^{DM}_{ij}={\mathsf{F}_{ij}}D_{i} D_{j}}$, 
where the i,j indices denote the residual numbers, $D_i=\frac{1}{K\nu_i}$, $\nu_i$ is the 
observing frequency of the TOA, typically set as the central frequency of the observing
band, and K=2.41$\times$10$^{-16}$ Hz$^{-2}$cm$^{-3}$pc s$^{-1}$, is the dispersion constant.
 
The likelihood function is the probability that the data (TOAs), noted as $\boldsymbol{t}$, are fully
described by the timing model signal, $\boldsymbol{\tau(\epsilon})$, with parameters $\boldsymbol{\epsilon}$ and the 
stochastic noise. The latter is encoded in the residuals' total covariance matrix, 
\begin{center}
\begin{equation}
\label{eq:CovMat}
{\bf C} = {\bf C_{\textrm{w}}} + {\bf C_{\textrm{DM}}} + {\bf C_{\textrm{TN}}} .
\end{equation}
\end{center}

Following \cite{2009MNRAS.395.1005V}, 
and noting that the difference ${\bf t}-\boldsymbol{\tau(\epsilon})$ gives the 
timing residuals vector, we can write the 
likelihood function as:
\begin{center}
\begin{equation}
\label{eq:BayesLikFunc}
L = \frac{1}{\sqrt{(2\pi)^n |{\bf C}|}} e^{-\frac{1}{2}({\bf t}-\boldsymbol{\tau(\epsilon}))^{\textrm{{\bf T}}}{\bf C}^{-1}({\bf t}-\boldsymbol{\tau(\epsilon}))} .
\end{equation}
\end{center} 
After the noise properties are estimated, we produce the TN waveforms, 
which can be estimated from the data using the maximum likelihood (ML) value of 
its statistical model parameters, $A$ and $\gamma$. As shown 
in \cite{2014MNRAS.441.2831L}, the ML waveform, ${\rm \bf t}_{\rm TN}$, and 
its uncertainties, ${\rm \bf \sigma}_{\rm TN}$, are optimally estimated as
\begin{equation}
	{\rm \bf t}_{\rm TN}={\rm \bf C}_{\rm TN} {\rm \bf C}^{-1} {\rm \bf t}\ ,
	\label{eq:waveestimator}
\end{equation}
with uncertainties estimated as:
\begin{equation}
	{\rm \bf \sigma}_{\rm TN}={\rm \bf C}_{\rm TN} - {\rm \bf C}_{\rm TN} {\rm 
	\bf C}^{-1} {\rm \bf C}_{\rm TN} .
	\label{eq:wavesigma}
\end{equation}
The uncertainties are estimated as the standard deviation of the estimator. 
However, as noted in \cite{2014MNRAS.441.2831L}, since the components of TN waveforms 
are correlated, their interpretation in terms of uncertainties is meaningless, since this
is only valid under the assumption that the noise is uncorrelated. 
The uncertainties can therefore only be used as an indication 
of the variance of each point. 

We have performed the Bayesian inference analysis twice using different combination of pPDs.
The pPDs on the timing parameters are always uniform, 
centred around the value from the wLS fit of the timing model by 
\ttwo{} with a range of 10 to 20 times their 1-$\sigma$ \ttwo{} uncertainties. 
This range was chosen after testing verified that is sufficient for 
all timing parameters PPDs to converge. For the noise parameters,
the ranges are from 0 to 7 for spectral indices, $-20$ to $8$ for 
the logarithm of the amplitudes, $-10$ to $-3$ for the logarithm of the EQUADs 
and $0.3$ to $30$ for the EFACs.
For EQUADs, TN and DM variations amplitudes we used 
two different types of pPDs. The first is a uniform distribution in log space (log-uniform) 
and the second is a uniform distribution in linear space (uniform). 
Log-uniform pPDs assume that all orders of magnitude are
equally likely for the parameter value while for uniform pPDs, 
we assign the same probability for all values. 
The uninformative log-uniform pPDs will result in PPDs for the parameters that are the least affected by the 
pPD and therefore are what we consider as the parameter measurement. 
If no significant noise can be detected in the data, the PPDs are unconstrained and 
the distribution's upper limit is dependent on the lower limit of the pPD. 
Therefore, a separate analysis is required using uniform pPDs in order to obtain robust upper limits.
If the signal is strong and the result from a log-uniform-pPD analysis is a well-constrained PPD, 
then the change of the pPD should not affect the result significantly and the PPDs should be almost identical. 
As a result, we performed the analysis with 
the following combination of pPDs:\\
a) Uniform EQUAD pPDs and log-uniform pPDs for TN and DM variation amplitudes. 
This set of pPDs results in upper limits for EQUADs. As such, the solutions have the highest possible 
timing residuals uncertainties, resulting in weaker 
TN and DM variations detections. The TN and the DM variations are 
treated in the same way, giving no prior information that can favour the one over the other 
when multi-frequency data are not sufficient to de-couple them. In the absence of multi-frequency data 
one can therefore expect that their PPDs will not be well-constrained.\\
b) Uniform TN amplitude and log-uniform pPDs for EQUADs and DM variation amplitudes: The total white noise levels 
of these solutions are lower, since EQUAD PPDs can be flat if the data do not support them to be measurable. 
The use of uniform pPDs for the TN amplitude and log-uniform for the DM variations results in 
solutions in favour of the TN against the DM variations in the absence of multi-frequency data. 
This set of pPDs will provide the strictest upper limits on the TN amplitudes. We used the 
PPDs from this analysis to calculate the amplitude upper limits 
at the 95\% confidence level (C.L.).\\

\begin{table*}
\centering
\caption{Timing-noise characteristics of EPTA MSPs based on Bayesian inference for 
a single power-law model as described by Equation~\eqref{eq:RedNoiseFunc}.
The results are divided based on the quality of the posterior probability distributions (PPDs) 
as described in Section~\ref{subsec:BvsF}. We tabulate the maximum likelihood (ML) and median (med)  
values of the dimensionless amplitude, A, at reference frequency of 1~yr$^{-1}$ and the 
the spectral index, $\gamma$. For A, we also tabulate the 95\% confidence upper limits.
The 1-$\sigma$ uncertainties are calculated such that the 68\% of the area 
under the 1-dimensional marginalised PPS of the parameter is symmetrically distributed around the median.
As described in Section~\ref{subsec:BvsF}, for unconstrained PPDs we only consider the upper-limits 
analysis results.}
\begin{tabular}{c c c c c c c}
\hline\hline
PSR  & log(A$_{\textrm{ML}}$) & log(A$_{\textrm{med}}$) & log(A$^{\textrm{95\%}}_{\textrm{UL}}$) &  $\gamma_{_\textrm{ML}}$ &   $\gamma_{_\textrm{med}}$ \\
J-Name & & & & & & \\
\hline\hline
& & &  Well-constrained PPDs \\
\hline
J0621+1002 & $-$12.029 & $-12.07^{+0.06}_{-0.06}$ & $-11.9$ & 2.5 & 2.4$^{+0.3}_{-0.2}$  \\[0.7ex]
J1012+5307 & $-$13.20 & $-13.09^{+0.07}_{-0.07}$ & $-12.94$ & 1.7 & 1.7$^{+0.3}_{-0.2}$   \\[0.7ex]
J1022+1001 & $-$13.2 & $-13.0^{+0.1}_{-0.2}$ & $-12.8$ & $2.2$ & 1.6$^{+0.4}_{-0.4}$  \\[0.7ex]
J1600$-$3053 & $-$13.35 & $-13.28^{+0.06}_{-0.06}$ & $-13.11$ & 1.2 & 1.7$^{+0.3}_{-0.2}$ &  \\[0.7ex]
J1713+0747 & $-$14.7 & $-15.2^{+0.5}_{-0.5}$ & $-13.8$ & 4.8 & 5.4$^{+0.9}_{-1.0}$ &  \\[0.7ex]
J1744$-$1134 & $-$13.7 & $-13.8^{+0.2}_{-0.3}$ & $-13.3$ & 2.2 & 2.7$^{+0.7}_{-0.6}$   \\[0.7ex]
J1857+0943 & $-$13.3 &  $-13.3^{+0.2}_{-0.3}$ & $-12.9$ & 2.6 & 2.4$^{+0.7}_{-0.6}$   \\[0.7ex]
J1939+2134 & $-$14.2 & $-14.5^{+0.3}_{-0.3}$ & $-13.7$ & 5.9 & 6.2$^{+0.5}_{-0.6}$  \\
\hline
& & & Semi-constrained PPDs \\
\hline
J0030+0451 & $-14.9$ & $-14.9^{+0.8}_{-2.1}$ & $-13.0$ & 6.3 & $5.2^{+1.2}_{-2.1}$  \\[0.7ex]
J0218+4232 & $-13.1$ & $-14.1^{+1.0}_{-1.7}$ & $-12.4$ & 2.7 & $3.9^{+1.7}_{-1.6}$   \\[0.7ex]
J0610$-$2100 & $-18.7$ & $-16.0^{+2.9}_{-2.7}$ & $-12.4$ & 1.4 & $2.7^{+2.8}_{-2.1}$   \\[0.7ex]
J0613$-$0200 & $-13.7$ & $-14.4^{+0.7}_{-0.9}$ & $-13.0$ & 2.8 & $4.1^{+1.6}_{-1.5}$  \\[0.7ex]
J0751+1807 & $-18.8$ & $-15.9^{+2.6}_{-2.7}$ & $-12.9$ & 6.5 & $3.0^{+2.0}_{-1.4}$   \\[0.7ex]
J1024$-$0719 &$-14.0$ &$-16.3^{+2.1}_{-2.4}$ &  $-13.1$  & 5.3 & $3.9^{+2.0}_{-2.5}$   \\[0.7ex]
J1455$-$3330 & $-19.8$ & $-14.2^{+1.0}_{-3.7}$ & $-12.7$ & 0.8 & $3.6^{+1.9}_{-1.6}$   \\[0.7ex]
J1640+2224 & $-13.2$ & $-13.1^{+0.2}_{-3.4}$ & $-12.8$ & 0.01 & $0.4^{+1.7}_{-0.3}$   \\[0.7ex]
J1643$-$1224 & $-17.7$ & $-13.3^{+0.6}_{-2.4}$ & $-12.5$ & 1.8 & $1.7^{+0.9}_{-0.6}$   \\[0.7ex]
J1721$-$2457 & $-11.7$ & $-13.5^{+1.7}_{-4.5}$ & $-11.5$ & 1.1 & $1.9^{+2.7}_{-1.0}$   \\[0.7ex]
J1730$-$2304 & $-12.8$ & $-14.7^{+1.7}_{-3.6}$ & $-12.6$ & 1.8 & $2.9^{+1.9}_{-1.3}$   \\[0.7ex]
J1801$-$1417 & $-14.4$ & $-15.1^{+2.5}_{-3.4}$ & $-12.2$ & 6.3 & $3.3^{+2.2}_{-1.8}$   \\[0.7ex]
J1802$-$2124 & $-17.0$ & $-15.6^{+3.2}_{-3.0}$ & $-12.2$ & 4.5 & $2.3^{+2.9}_{-0.8}$   \\[0.7ex]
J1843$-$1113 & $-13.0$ & $-12.9^{+0.2}_{-3.3}$ & $-12.5$ & 0.6 & $1.5^{+3.1}_{-0.5}$   \\[0.7ex]
J1909$-$3744 & $-14.1$ & $-14.1^{+0.2}_{-1.9}$ & $-13.8$ & 2.4 & $2.3^{+1.0}_{-0.6}$   \\[0.7ex]
J1918$-$0642 & $-16.9$ & $-14.5^{+0.7}_{-0.5}$ & $-12.6$ & 1.7 & $5.4^{+1.1}_{-1.6}$   \\[0.7ex]
J2145$-$0750 &$-14.4$ &$-14.0^{+0.6}_{-0.8}$ & $-12.9$  &5.2 &$4.1^{+1.6}_{-1.3}$   \\
\hline
& & & {\small Unconstrained} PPDs \\
\hline
J0034$-$0534 & - & - &  $-12.3$  & - & -  \\
J0900$-$3144 & - & - &  $-12.7$ & - & -   \\
J1738+0333 & - & - &  $-12.7$  & - &  - \\
J1751$-$2857 & - & - &  $-12.4$  & - &  - \\
J1804$-$2717 & - & - &  $-12.3$  & - & -  \\
J1853+1303 & - & - &  $-12.4$ &  - & -  \\
J1910+1256 & - & - &  $-12.1$  & - &  - \\
J1911$-$1114 & - & -  &  $-12.1$ & - &  - \\
J1911+1347 & - & - &  $-12.9$  & - &  - \\
J1955+2908 & - & - &  $-12.1$  & - &  - \\
J2010$-$1323 & - & - &  $-12.8$  & - & -  \\
J2019+2425 & - & - & $-11.9$  & - &  - \\
J2033+1734 & - & - & $-12.0$  & - &  - \\
J2124$-$3358 & - & - &  $-12.8$  & - & -  \\
J2229+2643 & - & - &  $-12.7$  & - &  - \\
J2317+1439 & - & - &  $-13.1$  & - &  - \\
J2322+2057 & - & - & $-12.3$  & - &  - \\
\hline
\hline
\end{tabular}
\label{tab:Bayes}
\end{table*}

\begin{table*}
\centering
\caption{Timing-noise characteristics of EPTA MSPs based on power-spectral analysis for a single power-law 
model as described by Equation~\eqref{eq:RedNoiseFunc}. 
We tabulate the dimensionless amplitude, A, at reference frequency of 1yr$^{-1}$, the spectral index, $
\gamma$, and the white-noise power level, $S_{\textrm{W}}$, and their respective 1-$\sigma$ uncertainties. We 
also tabulate the pre-whitening level used (level$_{\textrm{pw}}$).
For the pulsars where the measurement of timing noise was not possible, we quote the 95\% confidence upper limits for the amplitude. The table is divided as Table~\ref{tab:Bayes} for easier comparison.}
\begin{tabular}{c c c c c c c c c c}
\hline\hline
& & Measured &  \\
\hline
PSR & log(A) &  $\gamma$ & log(S$_{\textrm{W}}$(yr$^3$)) & level$_{\textrm{pw}}$  \\
J-Name &  & &  &  \\
\hline
J0621+1002 & $-12.3\pm$0.1 & 2.8$\pm$0.6 & $-26.94\pm$0.04 & 1  \\
J1012+5307 & $-13.01\pm$0.07 & 1.7$\pm$0.3 & $-28.60\pm$0.02 & 1  \\
J1022+1001 & $-13.2\pm$0.2 & 2.0$\pm$0.6 &  $-27.97\pm$0.03 & 0  \\
J1600$-$3053 & $-13.6\pm$0.1 & 1.3$\pm$0.5 &  $-29.36\pm$0.05 & 0  \\
J1713+0747 & $-14.2\pm$0.2 & 4.9$\pm$0.6  & $-30.146\pm$0.02 & 2 \\
J1744$-$1134 & $-13.6\pm$0.2 & 3.0$\pm$0.6  & $-28.90\pm$0.03 & 1  \\
J1857+0943 & $-13.2\pm$0.2 & 2.3$\pm$0.7  & $-27.97\pm$0.04 & 1  \\
J1939+2134   &  $-14.3\pm$0.1 & 6.7$\pm$0.5  & $-30.27\pm0.02$ & 2 \\
\hline
J0030+0451 &  $-13.2\pm0.4$ & $4.5\pm1.0$ &  $-27.78\pm$0.03 & 2 \\
J0218+4232 & $-12.6\pm$0.2 & 2.3$\pm$0.6 &  $-26.69\pm$0.03 & 0  \\
J0610$-$2100 & -13.6$\pm$0.1 & 2.1$\pm$0.6  & $-29.62\pm$0.03 & 0  \\
J0613$-$0200 & -14.9$\pm$0.9 & 5.2$\pm$1.8  & $-28.45\pm$0.03 & 0  \\
J0751+1807 & -14.3$\pm$0.7 & 5.2$\pm$1.6  & $-27.86\pm$0.03 & 1  \\
J1024-0719 &  $-13.0\pm$0.1 & 4.1$\pm$0.5 & $-28.15\pm$0.03 & 2 \\
J1455$-$3330 & $-13.4\pm$0.4 & 3.5$\pm$1.2 &  $-27.59\pm$0.03 & 0  \\
J1640+2224 & $-13.0\pm$0.1 & 1.4$\pm$0.4 &  $-27.96\pm$0.05 & 0 \\
J1643$-$1224 & $-13.2\pm$0.1 & 3.5$\pm$0.4  & $-28.25\pm$0.03 & 0  \\
J1721$-$2457 & $-12.3\pm$0.3 & 2.7$\pm$0.8  & $-26.01\pm$0.09 & 0  \\
J1730$-$2304 & $-12.8\pm$0.2 & 1.7$\pm$0.5  & $-27.31\pm$0.06 & 0 \\
J1801$-$1417 & $-13.3\pm$0.3 & 2.4$\pm$1.1  & $-28.41\pm$0.10 & 0  \\
J1802$-$2124 & $-12.8\pm$0.2 & 2.9$\pm$0.7  & $-27.93\pm$0.04 & 0  \\
J1843$-$1113 & $-12.8\pm$0.1 & 3.0$\pm$0.6  & $-27.93\pm$0.05 & 1  \\
J1909$-$3744 & $-14.5\pm$0.7 & 1.6$\pm$1.7  & $-30.05\pm$0.04 & 0  \\
J1918$-$0642 & $-13.0\pm$0.2 & 2.8$\pm$0.8  & $-27.72\pm$0.05 & 1  \\
J2145-0750 & $-13.7\pm$0.3 & 3.5$\pm$0.7  & $-28.36\pm0.03$ & 0\\
\hline
& & Upper Limits & \\
\hline
PSR &   log(A$^{\textrm{95\%}}_{\textrm{UL}}$) & &log(S$_{\textrm{W}}$(yr$^3$)) & level$_{\textrm{pw}}$  \\
J-Name & & & \\
\hline
J0034$-$0534  &  $-12.4$& -  & $-27.02\pm0.05$ &  0   \\
J0900$-$3144 & $-12.8$ &   - & $-28.0\pm0.1$     & 0\\
J1738+0333     & $-12.6$ & - & $-27.36\pm0.04 $   &  0   \\
J1751$-$2857  &  $-12.1$& -  & $-27.3\pm0.6$     & 0  \\
J1804$-$2717  &  $-12.2$&  -   & $ -26.57\pm0.09 $   & 0  \\
J1853+1303     & $-12.7$ & - & $ -27.7\pm0.1 $     & 0   \\
J1910+1256     & $-12.6$ & - & $ -27.38\pm0.06 $     & 0  \\
J1911$-$1114   & $-12.2$ & - & $ -26.7\pm0.1$ & 0 \\
J1911+1347 & $-12.8$ & -  & $-27.88\pm$0.1 & 0  \\
J1955+2908     & $-12.1$ & - & $ -26.46\pm0.06 $      & 0   \\
J2010$-$1323  & $-12.9$ &  -  & $ -27.95\pm0.04 $    & 0  \\
J2019+2425     & $-12.0$ & - & $ -26.14\pm0.08 $     & 0   \\
J2033+1734     & $-12.0$ & - & $ -26.15\pm0.06 $     & 0   \\
J2124$-$3358      & $-12.8$ & - & $ -27.69\pm0.04 $      & 0   \\
J2229+2643   &  $-12.7$ & - & $-27.66\pm0.05$  & 1 \\
J2317+1439     & $-12.8$ & - & $ -27.678\pm0.03 $      & 0   \\
J2322+2057    & $-12.3$  & -& $ -26.78\pm0.05 $       & 0   \\
\hline 
\end{tabular} 
\label{tab:Freq}
\end{table*} 
 
\subsection[]{Noise Parameter Estimation Using Power-Spectral Analysis}
\label{sect:Freq}
Power-spectral analysis of pulsar timing data using 
standard discrete Fourier transforms 
is complicated by highly variable error bars, irregular 
sampling, data gaps (due
to difficulties in being granted telescope time at exact regular intervals but also due to loss of data from technical 
difficulties, weather conditions, telescope maintenance or from weak pulses on particular days due to 
unfavourable interstellar scintillation) and the presence of TN which has a steep red spectrum. 
Fourier transforms require equispaced data points. Interpolation of data points on regular grids 
introduces time-correlations in data points and the presence of strong TN introduces spectral leakage. 
In order to bypass such problems, \cite{2011MNRAS.418..561C} introduced an 
algorithm for pulsar-timing analysis in the presence of 
correlated noise which 
employs the use of generalised least-squares (GLS) analysis of the timing data using the covariance 
matrix of the residuals (as described in Section~\ref{sect:Bayes}). In brief, the covariance matrix of the residuals 
is used to perform a linear transformation that whitens both the residuals and the timing model. The 
transformation is based on the Cholesky decomposition of the covariance matrix. 

For this algorithm, initial estimates of the residuals covariance 
matrix are necessary, and are obtained using the Lomb-Scargle periodogram (LSP), 
which can calculate the power spectrum of 
irregularly sampled data. Spectral leakage in the presence of strong TN with steep power-law spectra is 
mitigated with pre-whitening using the difference filter. The difference pre-whitening filter of any order, $k$, 
can be described by $y_{w,k}=y_{w,k-1}(t_i)-y_{w,k-1}(t_{i-1})$, where $t_i$ is the i-th
sampling time and $y_{w,k}$ is the whitened residual of difference order $k$ 
($k=0$ corresponds to the original residuals). It was suggested to use the lowest order necessary 
to whiten the data enough to mitigate spectral leakage. Effectively, this filter is equivalent to multiplying the 
power spectrum by a filter (e.g. for first order difference, the filter is the square of the transfer function). After the 
spectrum is estimated using the pre-whitened data, one corrects the power spectrum by dividing 
it with the same filter, a process known as post-darkening. 
The low-frequency spectrum can be fitted with a power-law model 
leading to the first estimation of the covariance matrix. 
Through an iterative process, new estimates of the spectrum 
can be achieved by using LSP after whitening the data 
using the Cholesky decomposition of the covariance matrix.

\cite{2011MNRAS.418..561C} have demonstrated that the implementation of this method
allows better timing solutions with more robust timing parameters and uncertainty calculations. In particular 
the measured spin and spin-down of the pulsar show the largest improvements, 
since they have low-frequency signatures in the Fourier domain and 
correlate with TN. However, this method is not optimised to accurately estimate the 
TN properties through detailed fitting of a noise model to the power-spectrum. 
The algorithm described in \cite{2011MNRAS.418..561C}
focuses on obtaining a linear, unbiased estimator of the timing parameters. For this purpose, they 
demonstrate that using the GLS timing solutions using the covariance matrices 
of any TN models which whiten the data sufficiently
to remove spectral leakage, are statistically consistent. 
In this work, we extend the algorithms of  \cite{2011MNRAS.418..561C}, focusing on the 
precise evaluation of the power spectra and the power-law model parameters. 
To this end, we have developed an 
independent power spectral analysis and model fitting code. 

A fully frequentist analysis should include a white-noise and DM-correction analysis. 
However, in order to focus on comparing 
the methods with regards to the estimation of the TN properties, we use the ML EFAC and EQUAD values 
and subtract the ML DM variations waveforms derived from the Bayesian analysis.

Our spectral analysis code calculates a 
generalised LSP, i.e. it performs a wLS fit of sine and cosine pairs at each frequency. We follow 
an iterative procedure as follows: (1) We first use \ttwo{} to obtain the wLS 
post-fit residuals, while subtracting the ML DM variations 
signal estimated with the Bayesian methods described in Section~\ref{sect:Bayes}. 
(2) We calculate the spectrum of these 
residuals using a chi-squared minimisation fit on all frequency points. 
(3) \ttwo{} is re-run using the covariance matrix of the initial noise model to perform a GLS fit. 
(4) Finally, we re-run the spectral analysis code on the residuals from the GLS timing solution to 
update the TN model and repeat steps 3 and 4 until the solution converges. 
Typically, this required no more than one iteration. 

Our code implements a generalised LSP to account
for the timing residual uncertainties. Denoting each pair of time and residual as ($t_i,y_i$), the LSP 
is formed by fitting sine-cosine pairs of the form ${\hat y}(\omega_k,t_i)=a_k \cos(\omega_k t_i) + b_k 
\sin(\omega_k t_j)$
at all angular frequencies, $\omega_k=2\pi f_k$, with $f_k$ the frequency. 
The solution is obtained by minimising the chi-squared for each $\omega_k$, weighted by the 
summed uncertainties of the timing residuals as:

\begin{center}
\begin{equation}
\label{eq:FrequenLS1}
\chi_{k}^2=\sum\limits_{i} \Bigg(\frac{y_i-a \textrm{sin}(\omega t_i)-b \textrm{cos}(\omega t_i)}{ {\hat \sigma_i}}\Bigg)^2 .
\end{equation}
\end{center}
Once the LSP is calculated, noting the number of timing residuals as 
$N$, the spectral density is finally computed as:

\begin{center}
\begin{equation}
\label{eq:FrequenLS5}
S(f)=\frac{2|{\hat y}|^2 T}{N^2} .
\end{equation}
\end{center}

We examine whether spectral leakage is present following the same routine as in \cite{2011MNRAS.418..561C}. 
Visual inspection of the original spectrum allows to approximately 
define the frequency where the red component of the 
spectrum intersects the flat, white component. 
We apply a low-pass filter in time-domain 
to separate the high-frequency from the low-frequency residuals and calculate their individual spectra. 
The high-frequency spectrum should be consistent with the high-frequency 
part of the spectrum of the original data. If that is not the case, and instead the high-frequency spectrum is significantly
weaker, then leakage is important and we need to apply the pre-whitening filter. 
The code allows for any order of difference whitening. For this data set, 
we required only up to second order. We then proceed with calculating
the LSP as before and finally post-darken the spectrum before calculating the final spectral density.

We fit the power spectrum with the following function:
\begin{center}
\begin{equation}
\label{eq:SA_RedNoise}
S(f)=S_{0}\Big(\frac{f}{f_r}\Big)^{-\gamma}+S_{\textrm{W}},\ \ S_{0}=\frac{A^2}{12\pi^2} .
\end{equation}
\end{center} 
Here, $S_{\textrm{W}}$ is the spectral density of the high-frequency (white) component. The power-law 
description of the low-frequency component is equivalent to Eq.~\eqref{eq:RedNoiseFunc}, with 
$S_{0}$ the spectral density at reference frequency, $f_r$, which is set to 1yr$^{-1}$. A 
fit of only the low-frequency component is proven difficult; 
due to the steepness of the spectrum at low frequencies and moderate power of the TN in many MSPs,
only about five frequency points would be included in a pure power-law fit of only the red part of the spectrum. 
This leads to unstable fits without meaningful error estimations. 

The fit minimises the chi-squared, $\chi_{S}^{2}$. 
Chi-squared minimisation assumes that the spectrum is normally distributed. In principle, the power spectrum 
is a chi-squared distribution. However, in logarithmic space, the distribution is approximately Gaussian with 
variance of order unity. Therefore this is a good 
approximation if we fit the power-law model to 
the spectrum in logarithmic space. By doing so, we minimise the chi-squared defined as:
\begin{center}
\begin{equation}
\label{eq:SA_RedNoise_min}
\chi_S^2=\sum\limits_{i=1}^{N} \left\{ \log S_i - 
\log\left(S_{0}\left(\frac{f_i}{f_r}\right)^{-\gamma}+S_{\textrm{W}}\right) \right\}^2 ,
\end{equation}
\end{center} 
where $S_i$ and $f_i$ define the points of the spectrum for
each frequency bin, $i$, while simultaneously fitting for $S_0$, $\gamma$, 
and $S_{\textrm{W}}$. 
We first fit the spectrum while setting the uncertainties of the LSP points to one and then scale the
uncertainties to achieve a reduced chi-squared of unity. 

Once we obtain the values for the noise parameters, we construct the covariance matrix of the TN, ${\bf C_{\textrm{TN}}}$. 
The Fourier transform of the TN power-law model gives the covariance function, 
$c_{\textrm{TN}}(\tau)=\langle{\textrm t}_{\textrm TN,i} {\textrm t}_{\textrm TN,j} \rangle$. The i and j indices refer to the 
time epoch of the observation and $\tau=i-j$. The TN covariance matrix is then formed 
by the elements  $C_{\textrm{TN},ij}=c(\tau_{\textrm{TN},ij})$, where $\tau_{ij}= |i-j|$. 
Using the total covariance matrix (Eq.~\ref{eq:CovMat}), we 
then perform a \ttwo{} GLS fit on the TOAs, repeat the power-spectrum analysis and power-law fit to 
update the model parameter values and iterate these steps until we converge to a stable solution. 

For the cases where the spectra are white-noise dominated and no measurement of the TN parameters can be 
achieved on a 3-$\sigma$ level, we derived upper limits for the TN amplitude. The limits are at the 
95\% C.L. and are calculated as the 2-$\sigma$ upper limit of the white-noise level 
($S_{\textrm{W}}$ in Eq.~\eqref{eq:SA_RedNoise} and Table~\ref{tab:Freq}).

\section[]{Results}
\label{sect:results}
Table~\ref{tab:Bayes} summarises the results of the noise properties 
determined using \tn{}, while Table~\ref{tab:Freq},
summarises the results from the power spectral analysis. The reader can find online\footnote{http://www.epta.eu.org/aom/DR1noise.html}
the PPDs of  the TN properties from the Bayesian analysis, the power spectra and the TN waveforms from both 
methods.
In the rest of this Section, we first discuss the framework under which we compare the  
results from the two methods and then 
proceed with the comparison of the results in more detail.
We conclude this Section by presenting and discussing the results on the white noise parameters
from the Bayesian analysis.

\subsection{Comparing Bayesian and Frequentists Results}
\label{subsec:BvsF}
Bayesian analysis is based on the principle that we test a hypothesis (model) given the data and a pPD. 
The latter is essential in Bayesian inference and states our prior degree of 
confidence on what the PD of the parameter is. The inference results in the PPD, which is the  
updated probability distribution for the unknown parameter, based on the information provided by the data.
\begin{figure*}
\begin{center}$
\begin{array}{cc}
\includegraphics[width=9.1cm, angle=0]{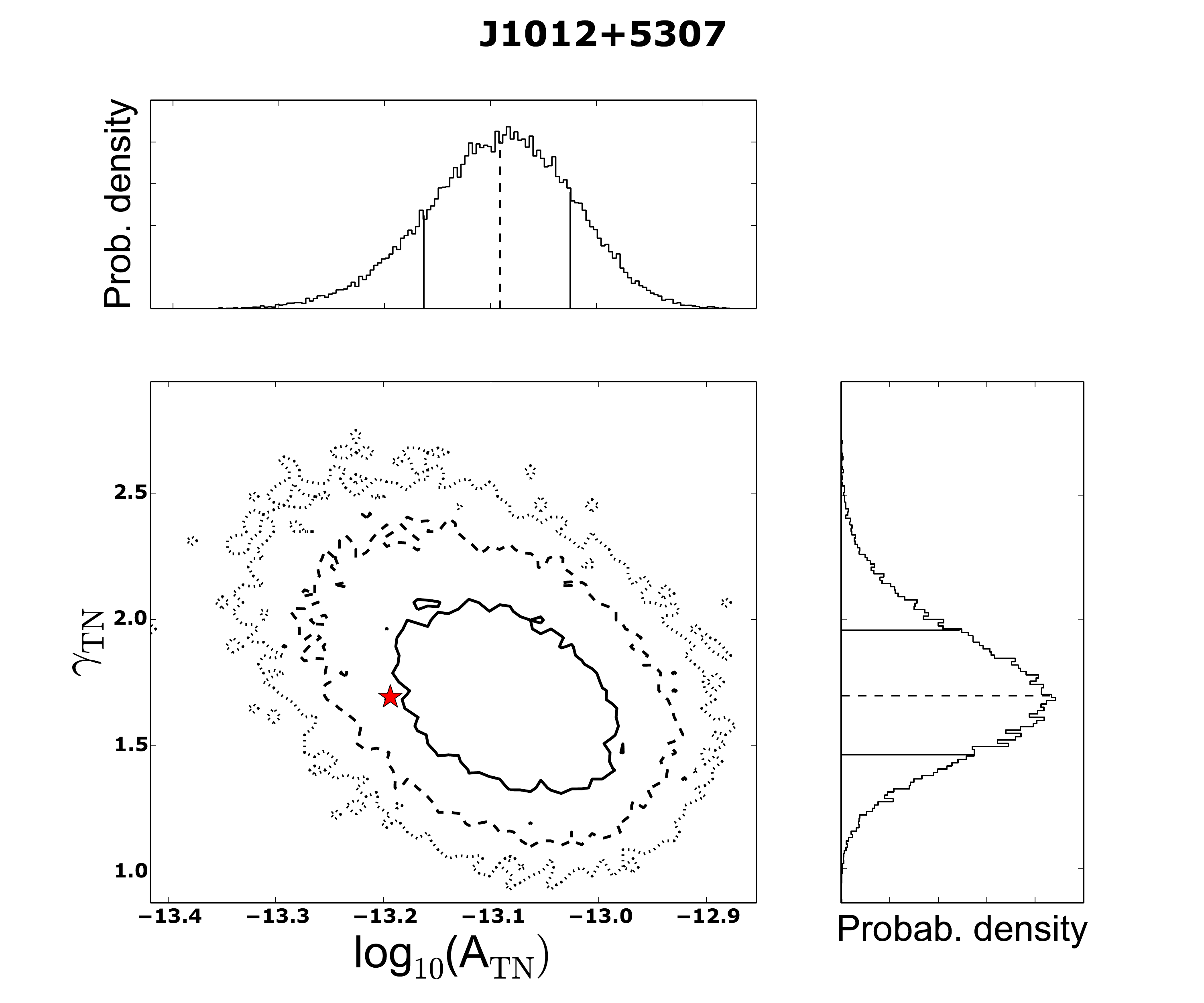} &
\includegraphics[width=9.1cm, angle=0]{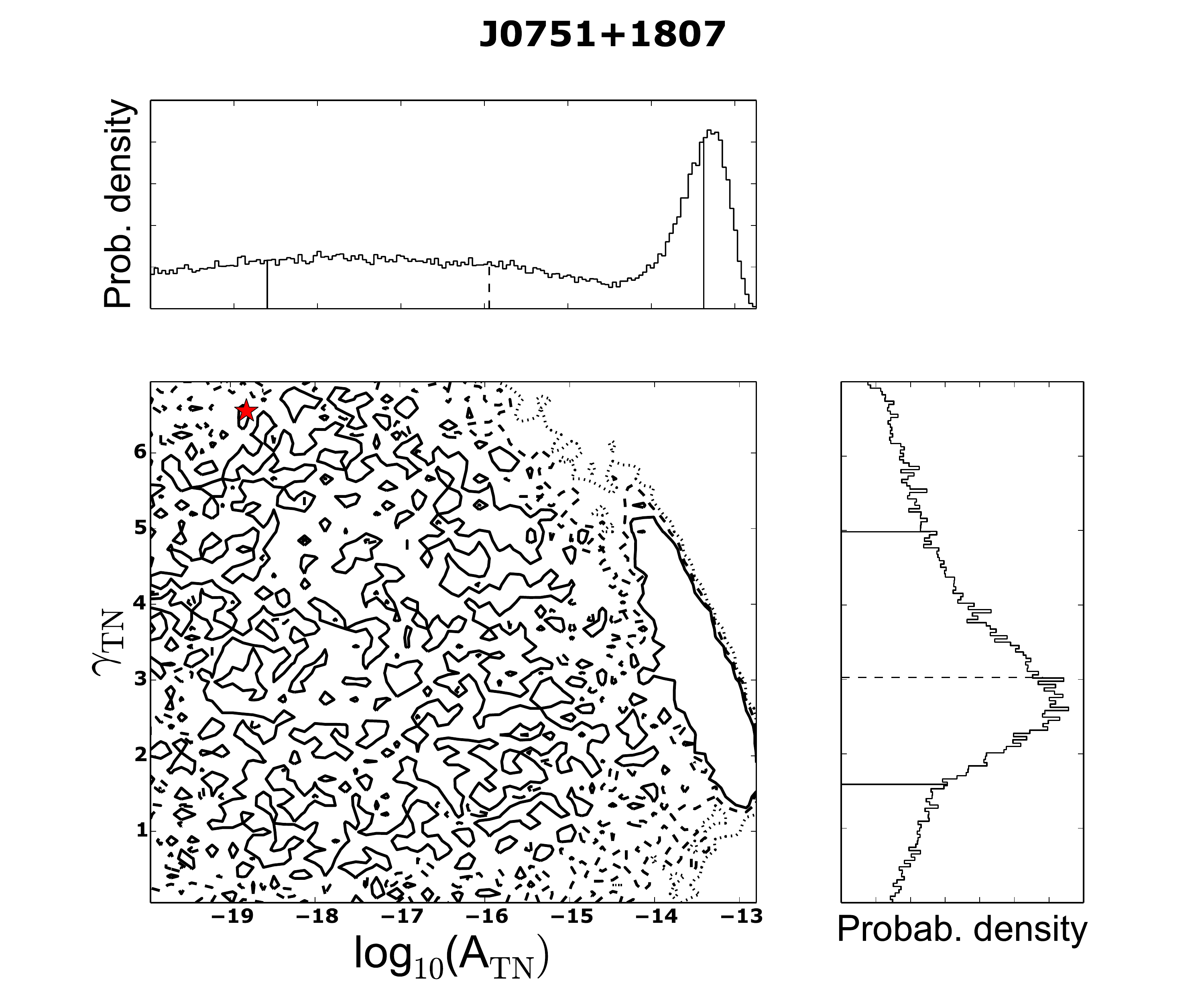} \\
\includegraphics[width=9.1cm, angle=0]{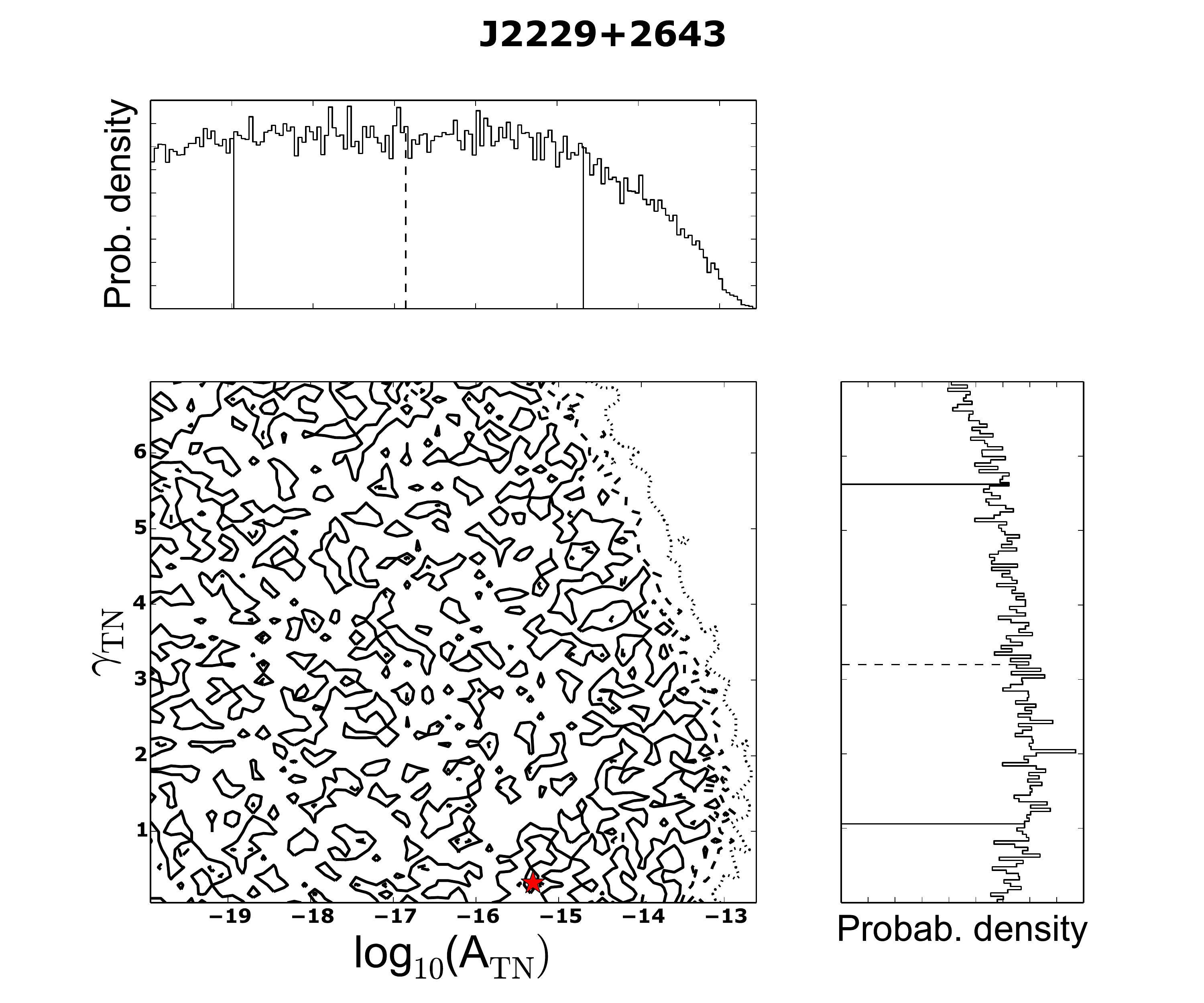} &
\includegraphics[width=9.1cm, angle=0]{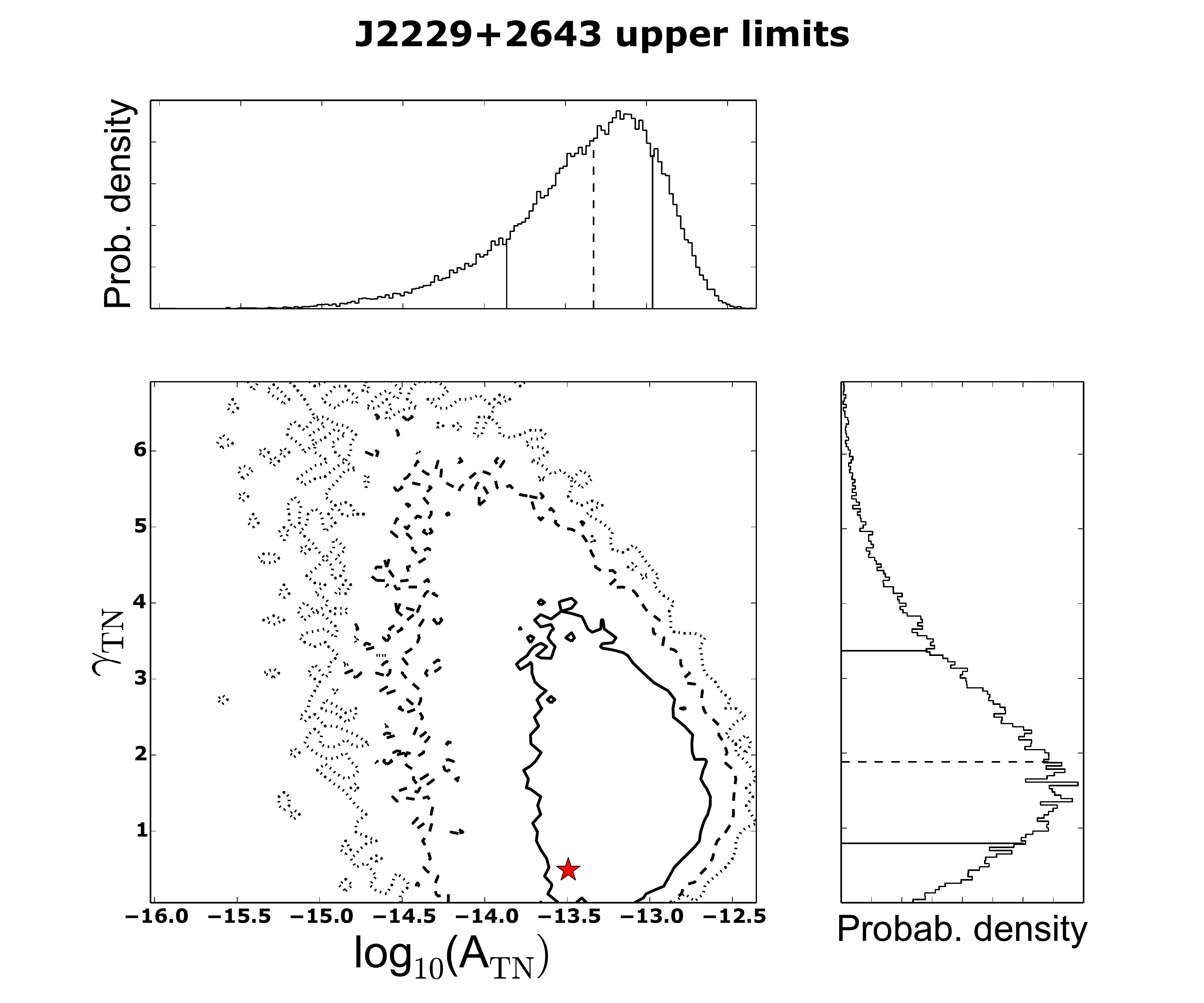} \\
\end{array}$
\caption{Two- and one-dimensional marginalised PPDs for the timing noise parameters of three pulsars: J1012+5307, J0751+1807,
and J2229+2643. 
In the two-dimensional distributions, the solid, dashed and dotted contours represent the 68\%, 
95\% and 99.7\% (1-, 2- and 3-$\sigma$) confidence intervals and the red star marks the maximum likelihood 
solution. The 1-dimensional distributions have the median and 
1-$\sigma$ uncertainties marked as dashed and solid lines respectively. 
For J2229+2643, the right figure shows the 
distribution of the noise parameters from the upper limits analysis. 
Note the different ranges on the amplitude axes.
See text in Section~\ref{subsec:BvsF} for discussion.} 
\label{fig:2d1d}
\end{center} 
\end{figure*}  
Bayesian inference also assigns the likelihood value 
for each model (i.e. for each set of values 
for all unknown parameters), providing a measure of how 
well the model describes the data. 
To evaluate the \tn{} results, we report in Table~\ref{tab:Bayes} 
the ML values of the TN parameters and 
the median value and 1-$\sigma$ uncertainties of the one dimensional marginalised PPDs. 
The uncertainties are calculated such that 68\% of the area 
under the distribution is symmetrically distributed around the median. The 
asymmetry of many PPDs will result in asymmetric error bars. 

We sort the PPDs in three categories, and show representative examples in Figure~\ref{fig:2d1d}. 
We name the first category of distributions ``well-constrained''; this represents cases where the data 
were sufficient to obtain good measurements of the noise parameters. As seen in Figure~\ref{fig:2d1d} 
for the case of PSR J1012+5307, the PPDs are well defined and very close to symmetric. As a result, 
the median values of the 1-dimensional PPDs coincide well with the ML solution. 
There are cases where the PPD of at least one of the TN parameters suffers from 
long tails due to strong covariances 
between unknown parameters (e.g. amplitude of TN 
and amplitude of DM variation noise in the absence of sufficient multi-frequency data). We refer to these 
distributions as ``semi-constrained''. As seen for the case of PSR J0751+1807 in Figure~\ref{fig:2d1d}, 
the two-dimensional distribution shows a main area of high probability as well as many smaller 
regions of local maxima. The tails in the one dimensional distribution of amplitude 
(which in general extend to $\pm \infty$), causes 
the median value to vary significantly from the ML model. Moreover, the large amount of 
area under the curve, along the tail, causes the uncertainties around the median to have large 
and very asymmetric values. Finally, when the data do not support 
any evidence of TN, the PPDs are flat. We refer to these as ``unconstrained''. As seen for the 
case of PSR J2229+2643 in Figure~\ref{fig:2d1d}, the reported 
median and ML values do not hold a strong significance. 
The only meaningful result to report in such cases is the upper limits for the amplitude, as seen in the 
bottom right panel of Figure~\ref{fig:2d1d}. 

Power-spectral analysis provides single-value resultsfrom the power-law model fit to the power spectrum. 
This fit is performed under the assumption of Gaussian statistics. 
As discussed above, in the case of power spectra, this is only an approximation. Finally, 
the fit is dependent on the estimation of the uncertainties of the power spectrum points, which was 
ensured to be properly calculated by pre-whitening the data when TN caused spectral 
leakage. 
\begin{figure*}
\begin{center}$
\begin{array}{cc}
\includegraphics[width=9.1cm, angle=0]{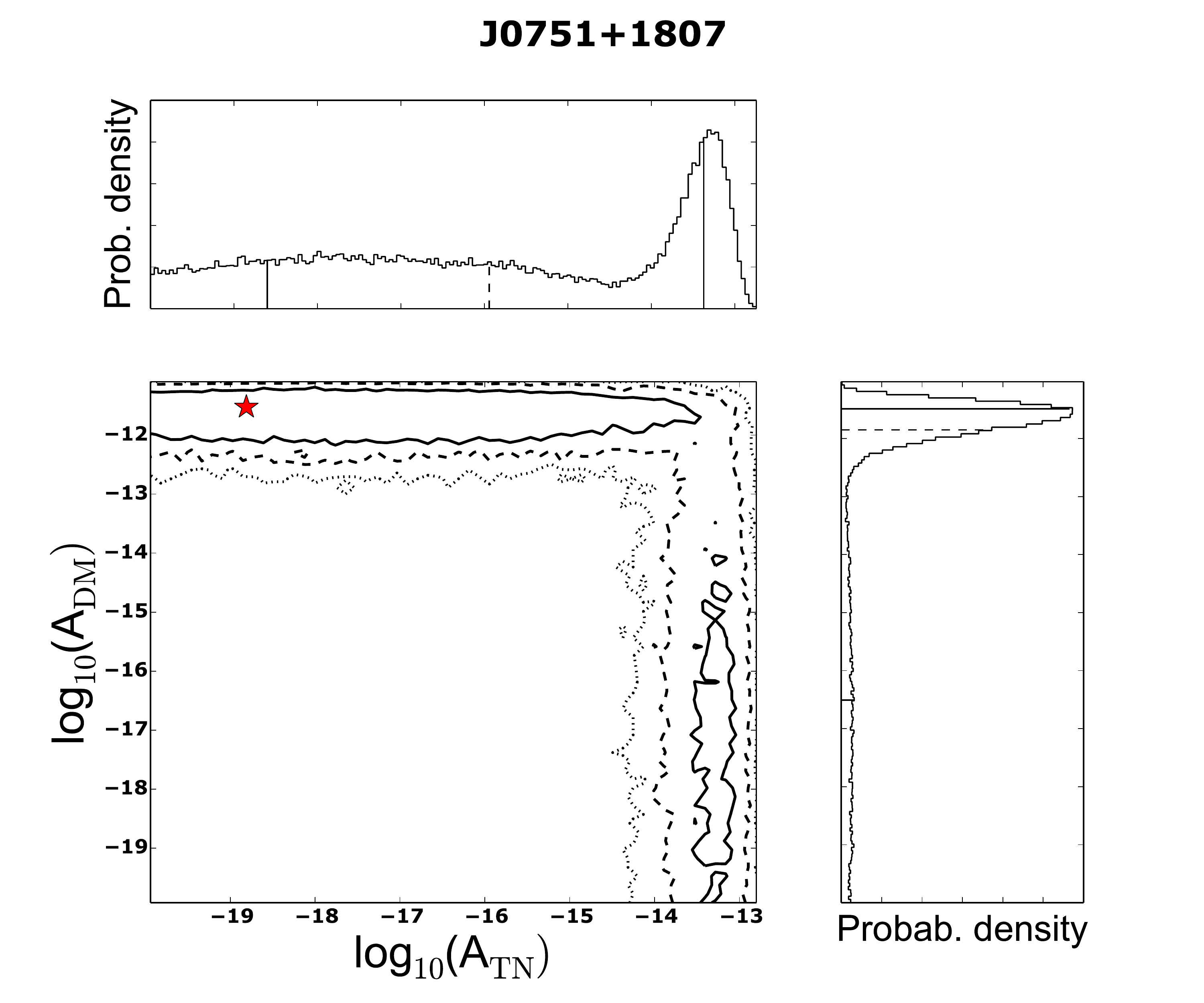} &
\includegraphics[width=9.1cm, angle=0]{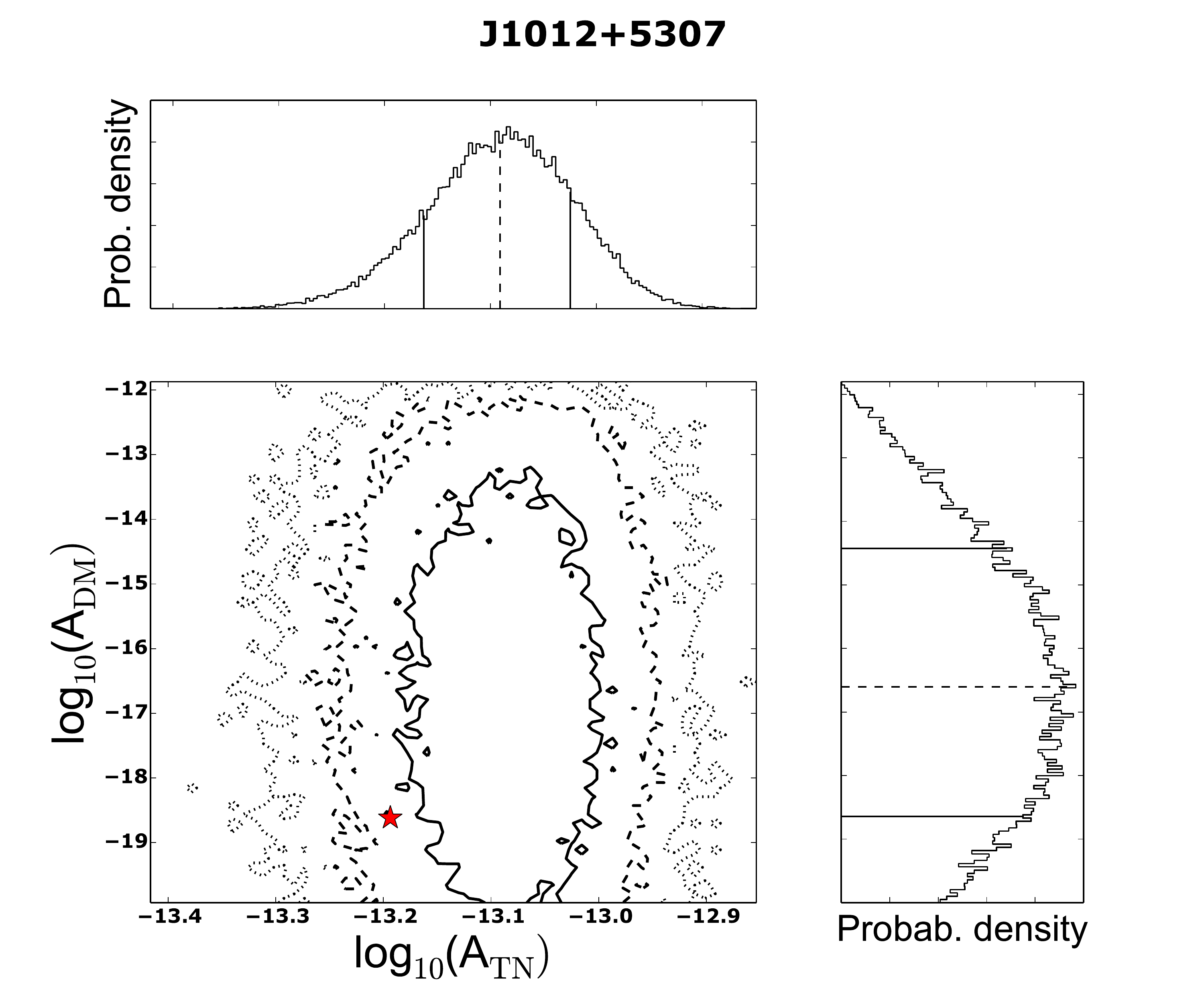} \\
\end{array}$
\caption{Two- and one-dimensional marginalised PPDs for the timing noise 
and DM variations amplitudes for J0751+1807 
and J1012+5307. In the two-dimensional distributions, the solid, dashed and dotted contours represent the 68\%, 
95\% and 99.7\% (1-, 2- and 3-$\sigma$) confidence intervals and the red star marks the maximum likelihood 
solution. The 1-dimensional distributions have the median and 
1-$\sigma$ uncertainties marked as dashed and solid lines respectively. Note the different ranges on the amplitude axes. See text in Section~\ref{subsec:BvsF} for discussion.} 
\label{fig:2d1d_RedDM}
\end{center} 
\end{figure*}
The comparison of the results derived with these two methods should also 
consider the effects of the Bayesian ML DM variations waveform subtraction 
from the residuals before performing the power-spectral analysis. In the case of semi-constrained 
PPDs, the amplitude parameters for the two TN and DM variations are naturally highly correlated. 
When this is the case, the ML parameter estimates are not as reliable, as the particular ML solution might correspond to either significant DM variations and no TN, or significant TN and no DM variations.
This can lead to over- or under-estimations 
of the DM variations which will lead to either part of the TN being subtracted as well or 
part of the DM signal leaking into the TN. 

As an example, we show in the left panel of 
Figure~\ref{fig:2d1d_RedDM} the 
two- and one-dimensional marginalised PPDs for the amplitudes of the timing noise and DM 
variations for PSR J0751+1807 (semi-constrained PPDs case). One can see the strong covariance between the 
two parameters. The data support that the TN amplitude is more likely to be very low 
(the TN tail has more probability than the DM tail), however, 
there is still a non-zero probability that the DM variations 
signal is weaker than the ML model suggests. For well-constrained PPDs, DM variations and TN 
are de-coupled, as seen in the right panel of Figure~\ref{fig:2d1d_RedDM} for the case of 
J1012+5307, and the DM ML waveform subtraction is more reliable. If the statistical 
assumptions of the Bayesian and frequentist analysis are valid, the results for the 
TN of pulsars with well-constrained PPDs should be consistent between the two methods.

\subsection{Timing-Noise Parameters}
\label{subsec:TNparams}
\begin{figure*}
\begin{center}$
\begin{array}{cc}
\includegraphics[width=9.1cm, angle=0]{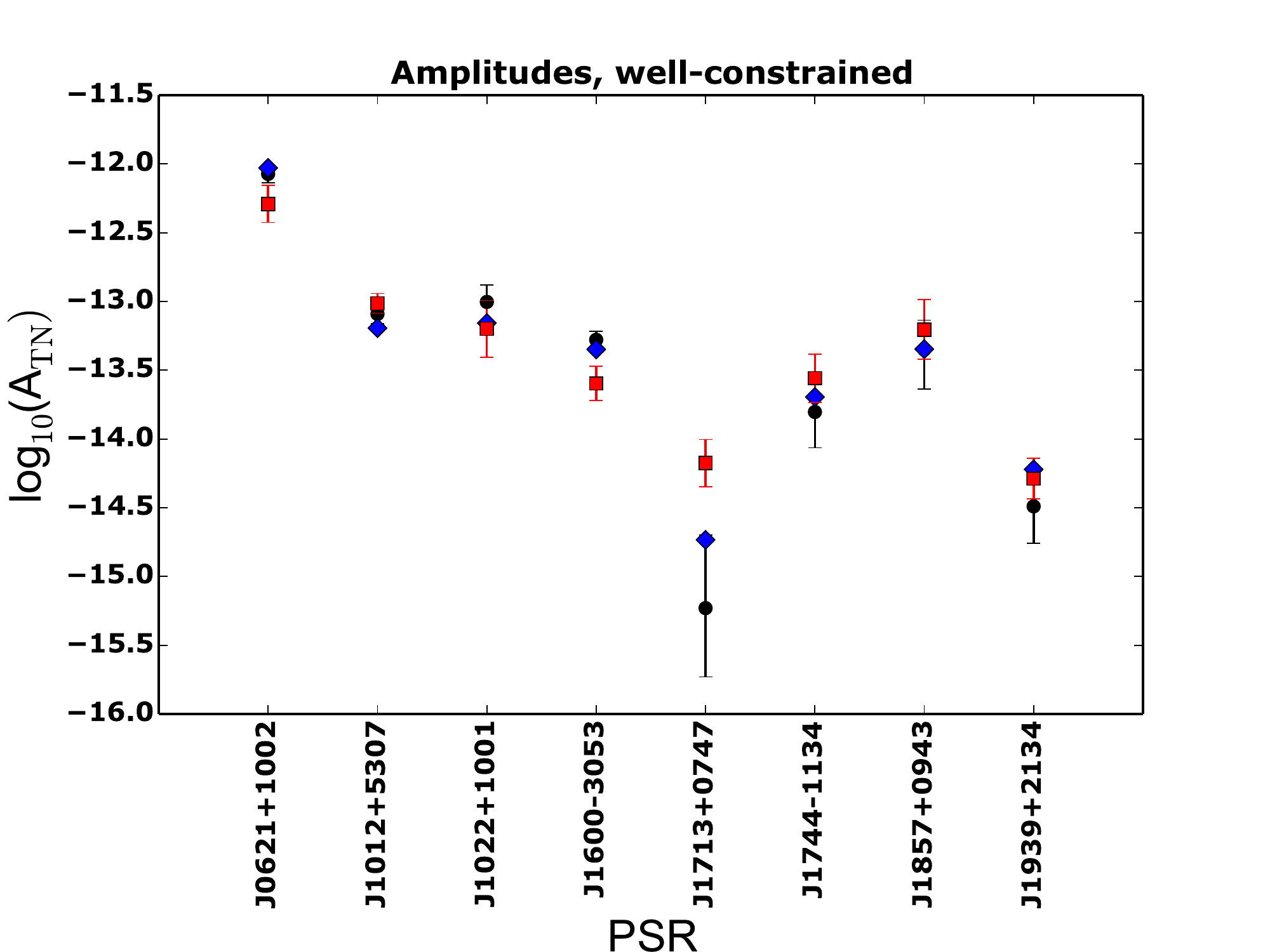} &
\includegraphics[width=9.1cm, angle=0]{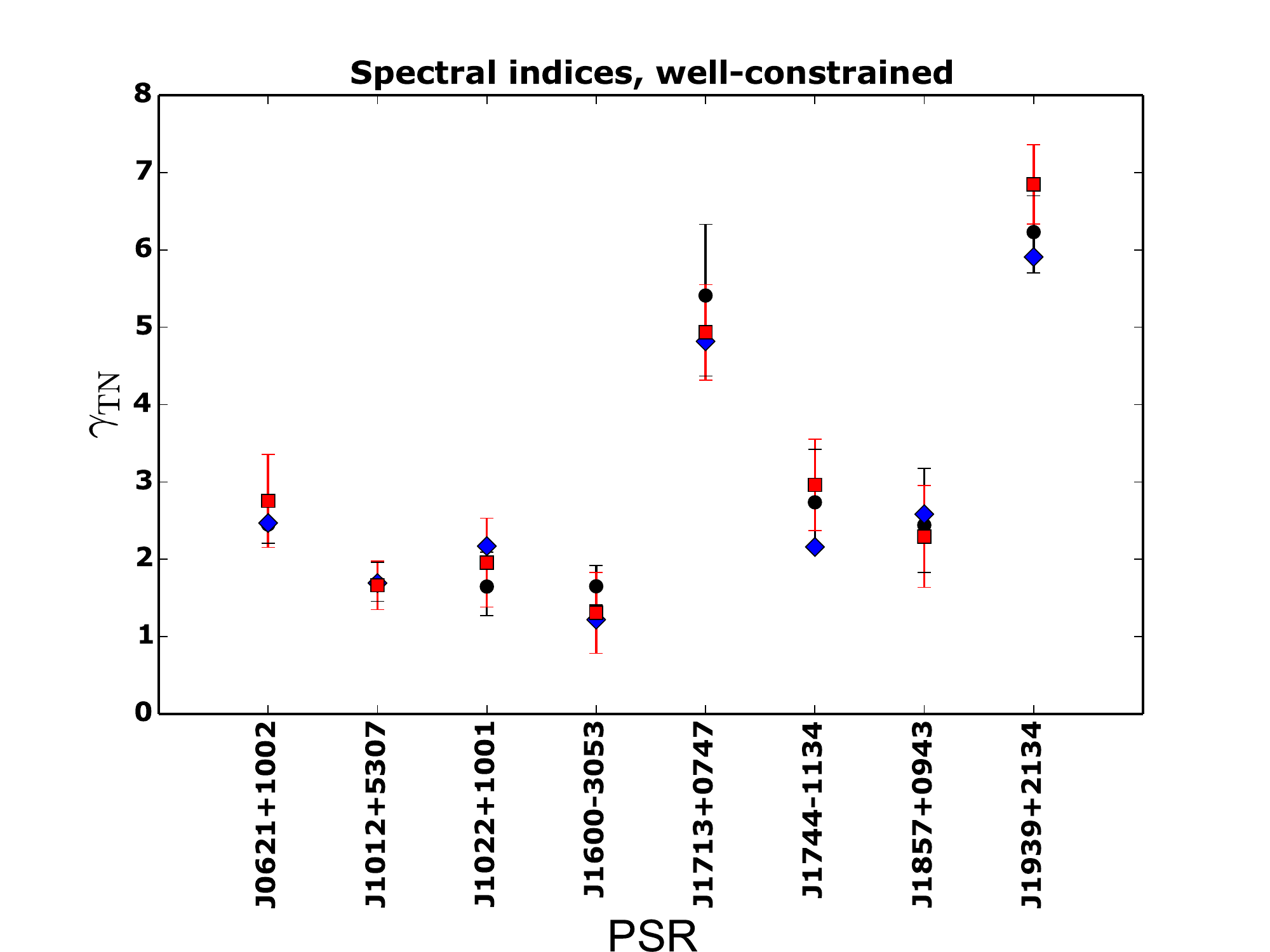} \\
\includegraphics[width=9.1cm, angle=0]{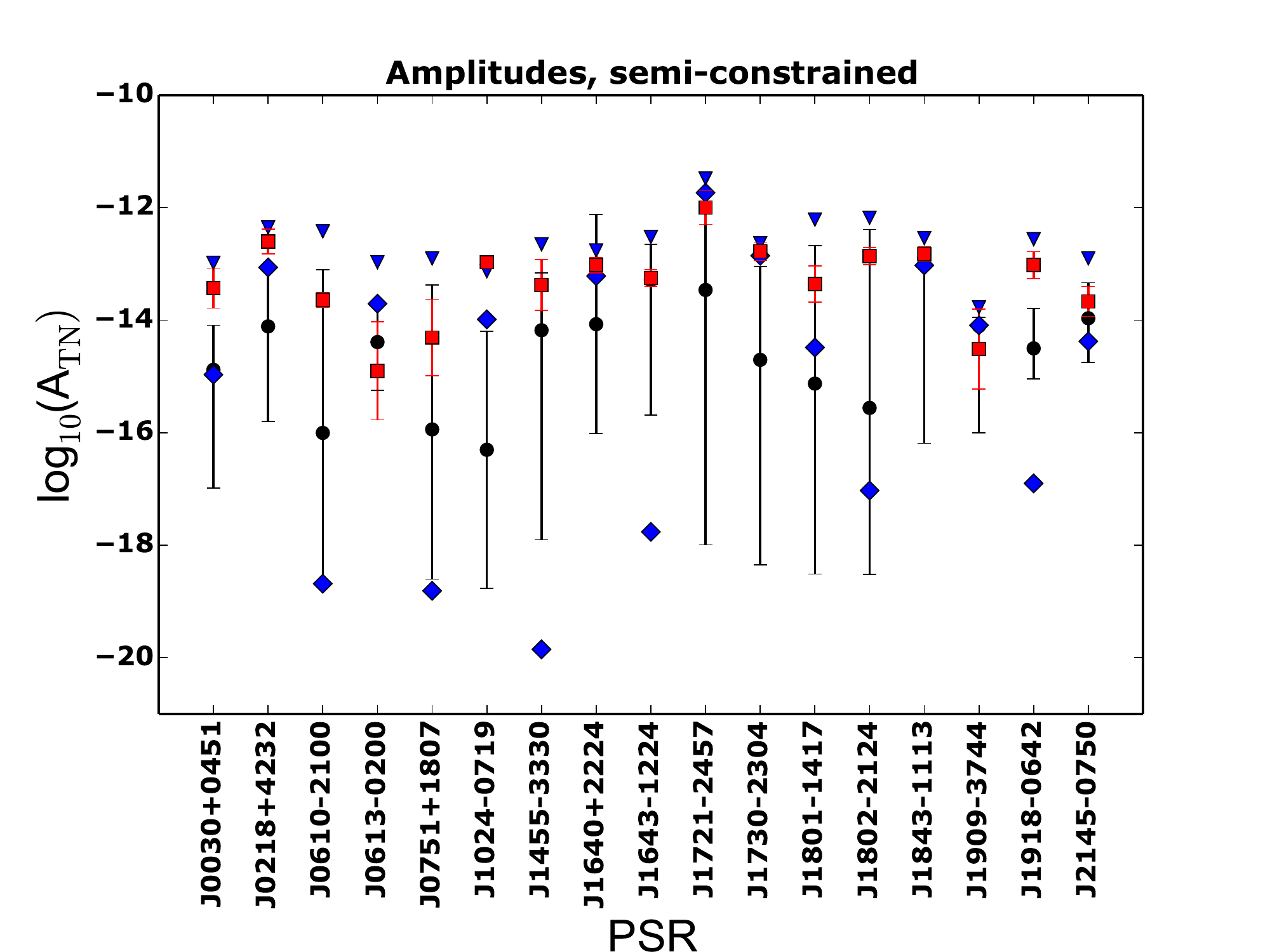} &
\includegraphics[width=9.1cm, angle=0]{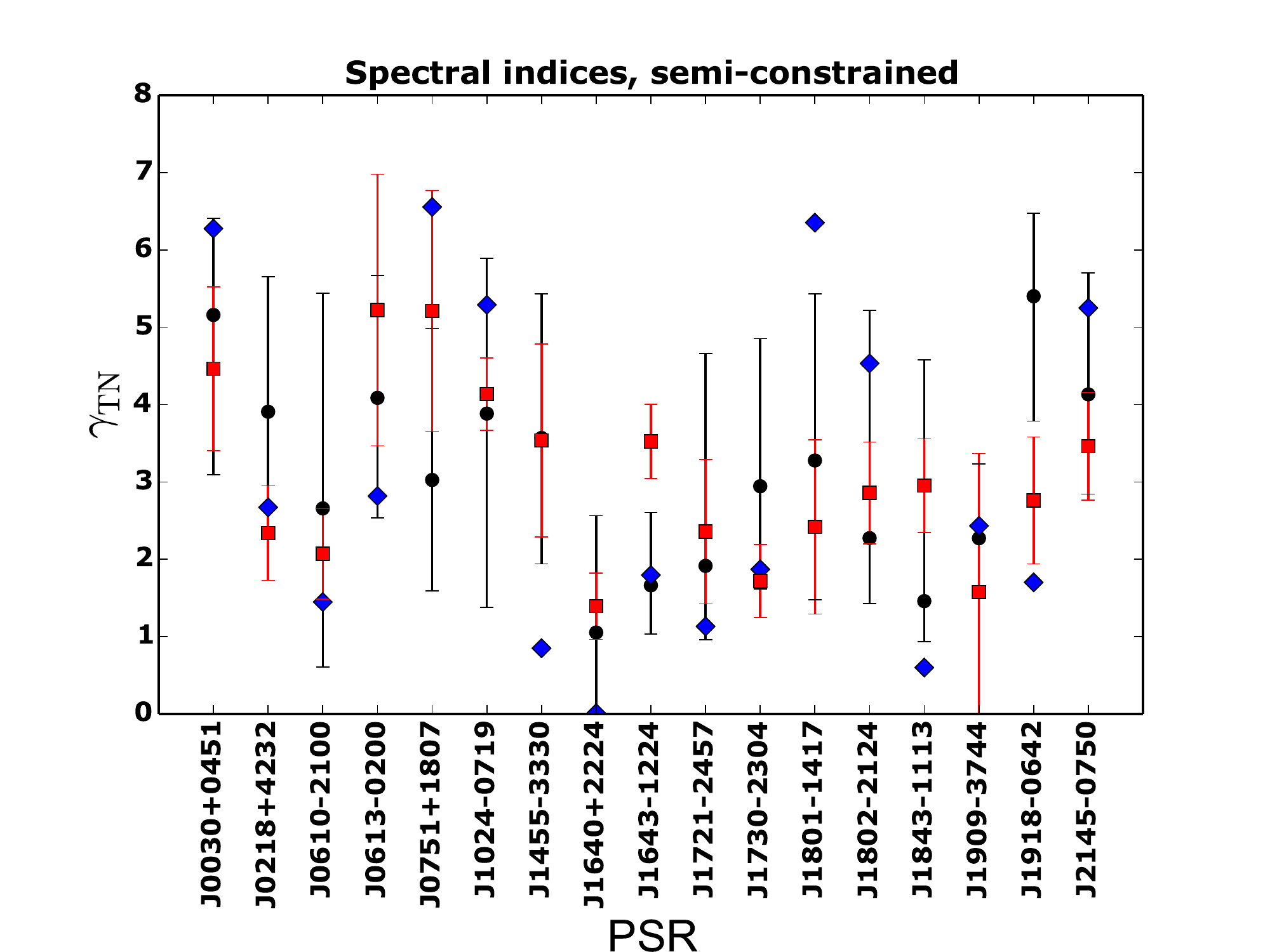} \\
\end{array}$
\includegraphics[width=9.1cm, angle=0]{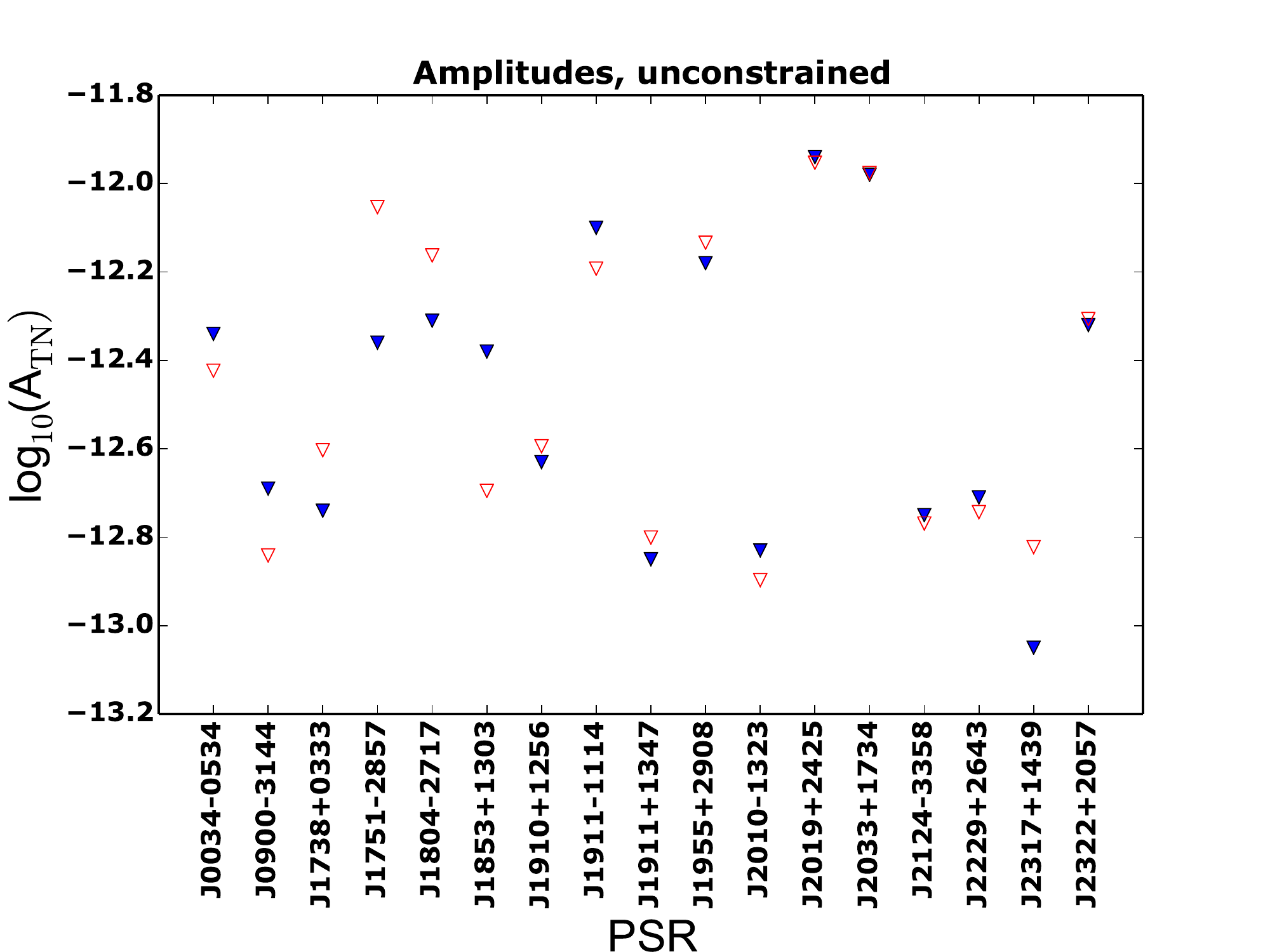}\\
\caption{Comparison of the timing noise parameters estimated with the Bayesian (blue, filled diamonds for the maximum
likelihood values, black, filled circles for the median values with 1-$\sigma$ error bars and blue, 
filled triangle for upper limits) 
and frequentist method (red, filled squares and red, open triangles for upper limits). {\it Top Row}: Results for the cases where the Bayesian analysis resulted in well-constrained posterior probability distributions for both parameters. 
{\it Middle Row}: Results for the cases where the Bayesian analysis resulted in 
semi-constrained posterior probability distributions for at least one of the parameters. 
{\it Bottom Row}:
Results for the cases where the Bayesian code resulted in unconstrained 
posterior probability distributions for at least one of the parameters.} 
\label{fig:BvsF01}
\end{center} 
\end{figure*}

Out of the 42 sources, the Bayesian analysis resulted in well-constrained PPDs for 
both the amplitude and the spectral index of the TN 
power-law model for eight sources. For these, the Bayesian ML and median values are 
always consistent at the 1-$\sigma$ level. The two methods are always consistent at the 
1-$\sigma$ level for the spectral index, while for the amplitude, three sources show deviations, though 
consistency remains at the  2-$\sigma$ level. (Figure~\ref{fig:BvsF01}, top row).

For 17 MSPs, the PPDs of at least one of the timing noise parameters is 
semi-constrained. The Bayesian ML and median values 
show inconsistencies at the 1-$\sigma$ level in four pulsars (Figure~\ref{fig:BvsF01}, middle row). 
The power-spectral analysis results are 
in agreement with the Bayesian median values. All Bayesian upper limits are 
in agreement with the rest of the results. We note that for PSR J1909$-$3744, 
we did not achieve a 3-$\sigma$ measurement for spectral index with the power-spectral analysis.

The rest of the sources, 17 in total, show flat, unconstrained PPDs. The bottom row of 
Figure~\ref{fig:BvsF01} shows the 95\% C.L. upper limits from the two methods. 
Given the low significance of the TN measurement in these cases, inconsistencies in the amplitudes 
do not have statistically significant effects on the timing solutions when using the total 
covariance matrix to perform GLS timing solutions.

The agreement between the two methods for the sources with 
statistically significant TN measurements, 
supports the confidence in the methods and the results. 
When covariances between noise properties cannot be decoupled by the data, 
the interpretation of the results requires more attention. For this 
reason, we propose that cross-checks of the results with various methods should become common practice. 

\subsection{White-Noise Parameters}
\label{subsec:BvsF2}
Radiometer noise estimation is typically robust when the pulse has a medium to high
signal-to-noise ratio (S/N) \citep{1992RSPTA.341..117T}, 
so EFACs are expected to be close to unity for most observing systems. 
The EQUADs results indicate for which observing systems there may be 
additional scatter in the residuals from physical processes related to the 
pulsars (e.g. pulse phase jitter) or RFI. 

\begin{figure}
\begin{center}
\includegraphics[width=9.4cm, angle=0]{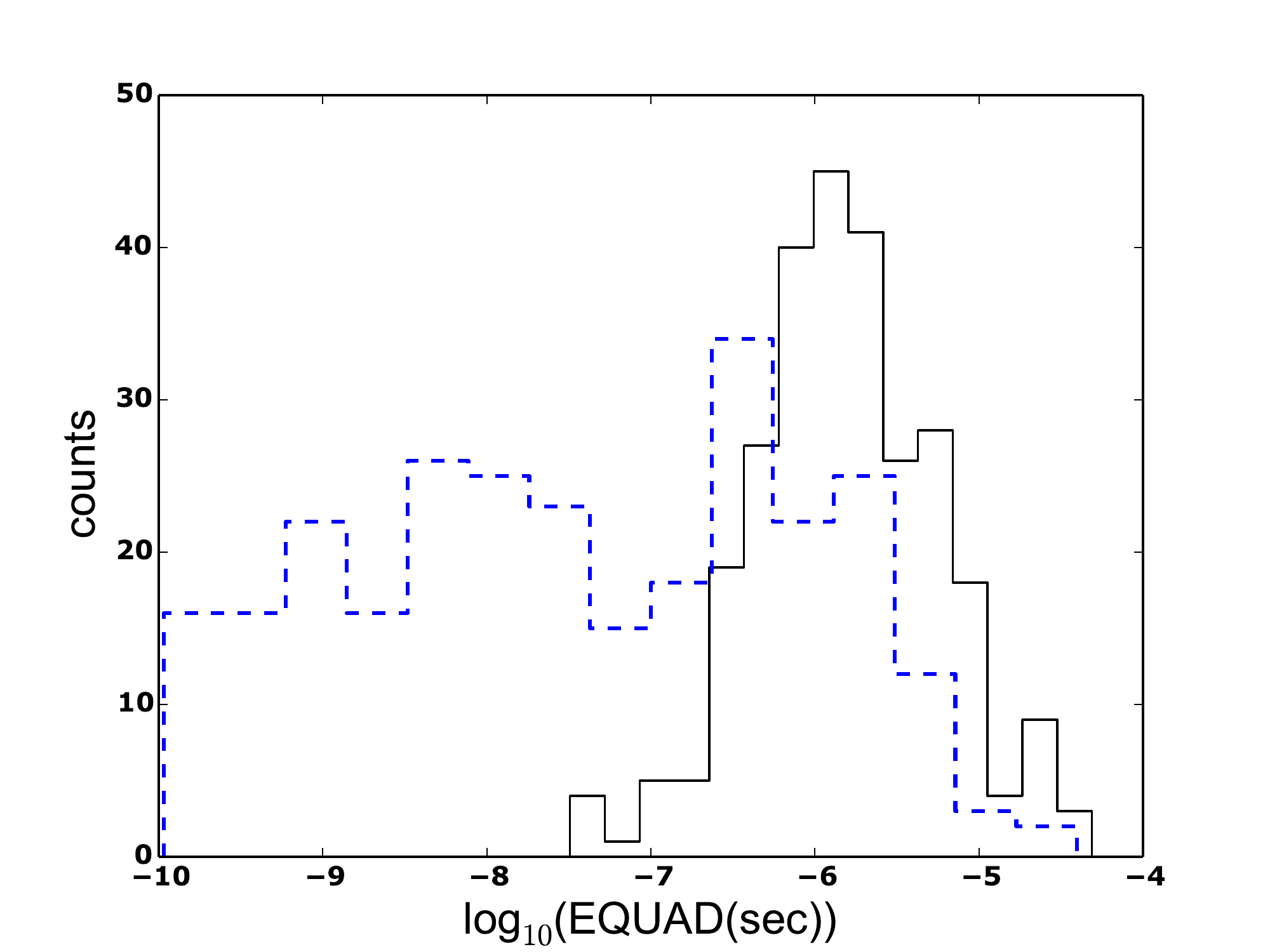}
\caption{Distribution of EFAC values for all MSPs and observing systems. 
The black, solid line refers to the results of the Bayesian
analysis for which the EQUAD priors are set to uniform to get 
their upper limit values, while the blue, dashed line is for the analysis
were EQUAD priors are uninformative log-uniform.} 
\label{fig:efacsHis}
\end{center} 
\end{figure}

Figure~\ref{fig:efacsHis} shows the distribution
of the ML EFAC values. As expected, the distribution strongly peaks around unity. 
A few systems show EFAC values up to $\sim$5. These are typically high-frequency observations 
with very weakly detected pulses. The cases where EFACs take values significantly lower than one are 
either due to strong overestimation of the uncertainties or when a system's EFAC and EQUAD 
are highly correlated.

\begin{figure}
\begin{center}
\includegraphics[width=9.4cm, angle=0]{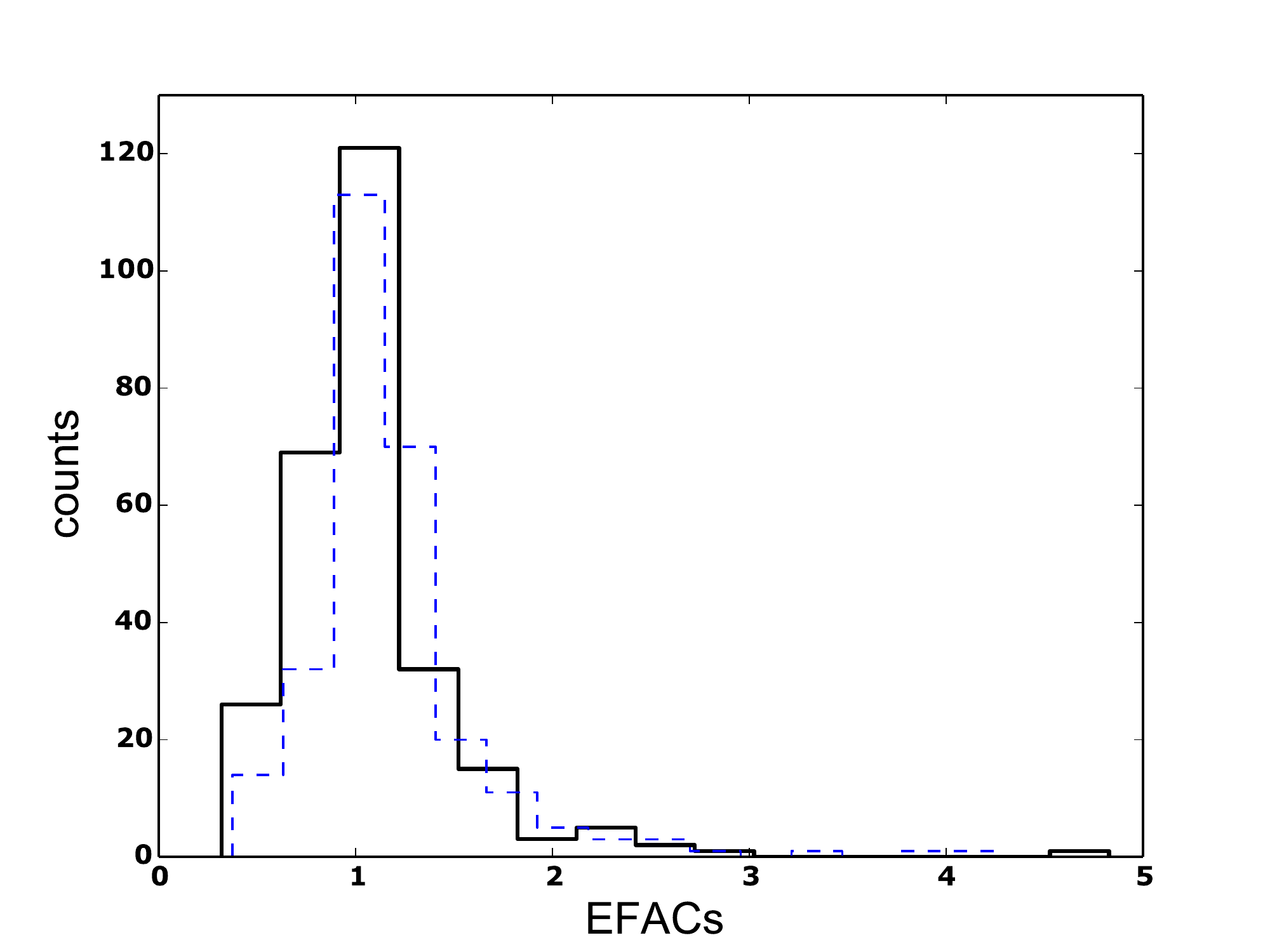}
\caption{Distribution of EQUAD values for all MSPs and observing systems. 
The solid line refers to the results of the Bayesian
analysis for which the EQUAD priors are set to uniform to get 
their upper limit values, while the dashed line is for the analysis
were EQUAD priors are uninformative log-uniform.}
\label{fig:equadsHis}
\end{center} 
\end{figure}

We examine in a similar way the distribution of EQUAD values. 
Figure~\ref{fig:equadsHis} shows the distribution of the measured ML EQUAD values from the analysis
using log-uniform EQUAD pPDs, and the distribution of their upper limits.
As expected, in the vast majority of cases, 
the EQUADs are much below the TOA precision, which typically ranges
from 0.5-10$\mu$s (D15). 

We have examined the EQUAD PPDs from the analysis with log-uniform pPDs to 
determine the cases where 
EQUADs have well-constrained PPDs and therefore show measurable EQUADs. 
For some of these cases, this could reflect signs of jitter noise present in the data.  
We list these pulsars and observing systems in Table~\ref{tab:equadsPDs}. 
We note that there are 49 cases where the EQUAD PPDs are semi-constrained and significantly covariant 
with EFACs, and therefore cannot be considered as significant EQUAD measurements. 
From Table~\ref{tab:equadsPDs} we can see that the vast majority of EQUADs come from
L-band systems, which typically have the most sensitive data. 
For each pulsar there are usually only one or two systems with clear EQUAD measurement with the exception 
of PSR J1022+1001. This source is known to require a high level of polarimetric calibration \citep{2013ApJS..204...13V} 
and to show phase jitter noise \citep{2015MNRAS.449.1158L}. Only part of the NRT data were calibrated and this 
may explain the high levels of EQUADs in this source. We stress once again, that more detail 
investigation is required to comment on the origin of the EQUAD measurements. It is likely 
that EQUADs could reflect additional scatter in the residuals from instrumental instabilities or 
analysis systematics, which could explain the EQUAD measurements in systems where the 
TOA precision is too low to expect any measurements of pulse jitter noise (as in the case e.g. of 
PSR J2033+1734, see Table~\ref{tab:equadsPDs}.)

\begin{table}
\begin{center}
\caption{List of the pulsars and observing systems for which we have well-constrained posterior probability distributions for the EQUADs. The last column shows the EQUADs maximum likelihood 
values from a Bayesian analysis with log-uniform EQUAD prior distribution. The telescope and 
backend acronyms are as 
introduced in Section~\ref{sect:Dataset}. }
\begin{tabular}{c c c c c}
\hline
\hline
PSR & Telesc. & Backend & Freq.  & EQUAD$_{\textrm{ML}}$\\
J-Name & & & (MHz) & ($\mu$s) \\
\hline
J0751+1807 & EFF & EBPP &1360 & $5.0$ \\
\hline
J1012+5307 & EFF & EBPP &1360 & $3.4$ \\
\hline
J1022+1001 & JBO & DFB & 1520 & $1.4$ \\
 & NRT & BON &1400 & $1.3$ \\
 & EFF & EBPP &1410 & $3.9$ \\
 \hline
J1643$-$1224 & JBO & DFB &1520  & $2.5$\\
\hline
 J1744$-$1134 &JBO & DFB & 1520 &  $1.0$\\
\hline
J1857+0943 & NRT & BON & 1400 &  $0.9$\\
\hline
J1939+2134 & NRT & DDS & 1400 &  $0.3$\\
 & EFF & EBPP &1410 & $0.3$ \\
 \hline
J2033+1734 & NRT & BON &1400 & $25$\\
\hline
J2145$-$0750 & NRT & BON & 2000 &  $0.3$\\
 & JBO & DFB & 1520 & $0.9$  \\
\hline
\end{tabular}
\label{tab:equadsPDs}
\end{center}
\end{table}

\section{Timing noise from Individual Observing Systems}
\label{sect:systematics}
For MSPs which have large enough data span 
with overlapping data from various observing systems we 
examine whether part of the measured TN is present only in specific observing systems.
We perform the noise analysis on selected pulsars with data from one telescope removed at a time. 
For the Effelsberg data, this is more complicated for many MSPs 
where it is the only telescope with data in the first half of the
data set, so removing its data automatically means a loss of about half the data span.
We note that this 
test may not be feasible in some cases with this data set,
e.g. when a significant fraction of the residuals sensitivity to the TN is lost 
when removing a set of dominant, very precise data points. 
When the TN was absent after removing data from one 
telescope, we confirmed that the rest of the data would be sufficient to detect the noise by simulating realisations 
of the new data and performing the noise analysis after injecting TN with the measured properties. 

\begin{figure*}
\begin{center}$
\begin{array}{cc}
\includegraphics[width=9cm]{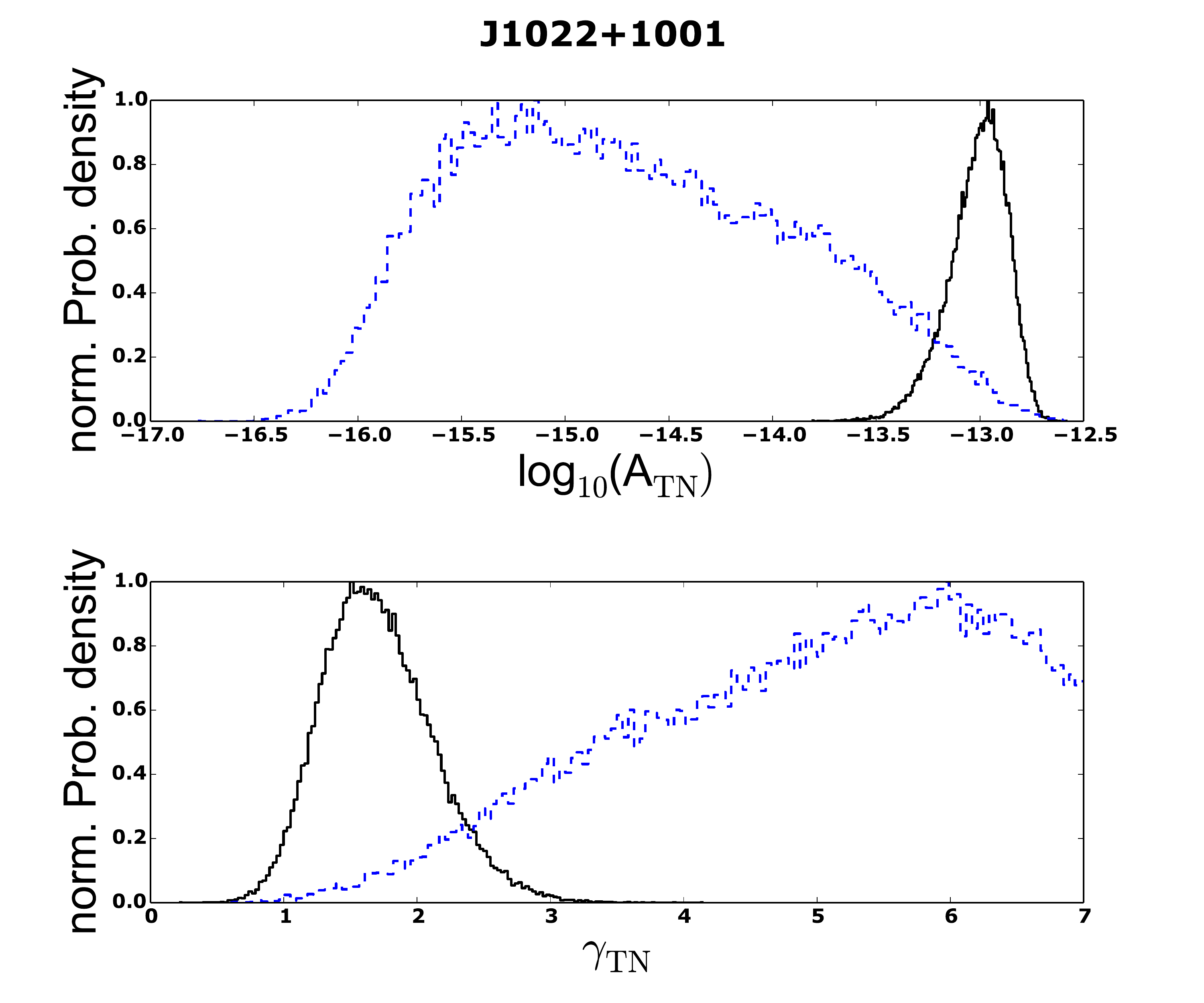} &
\includegraphics[width=9cm]{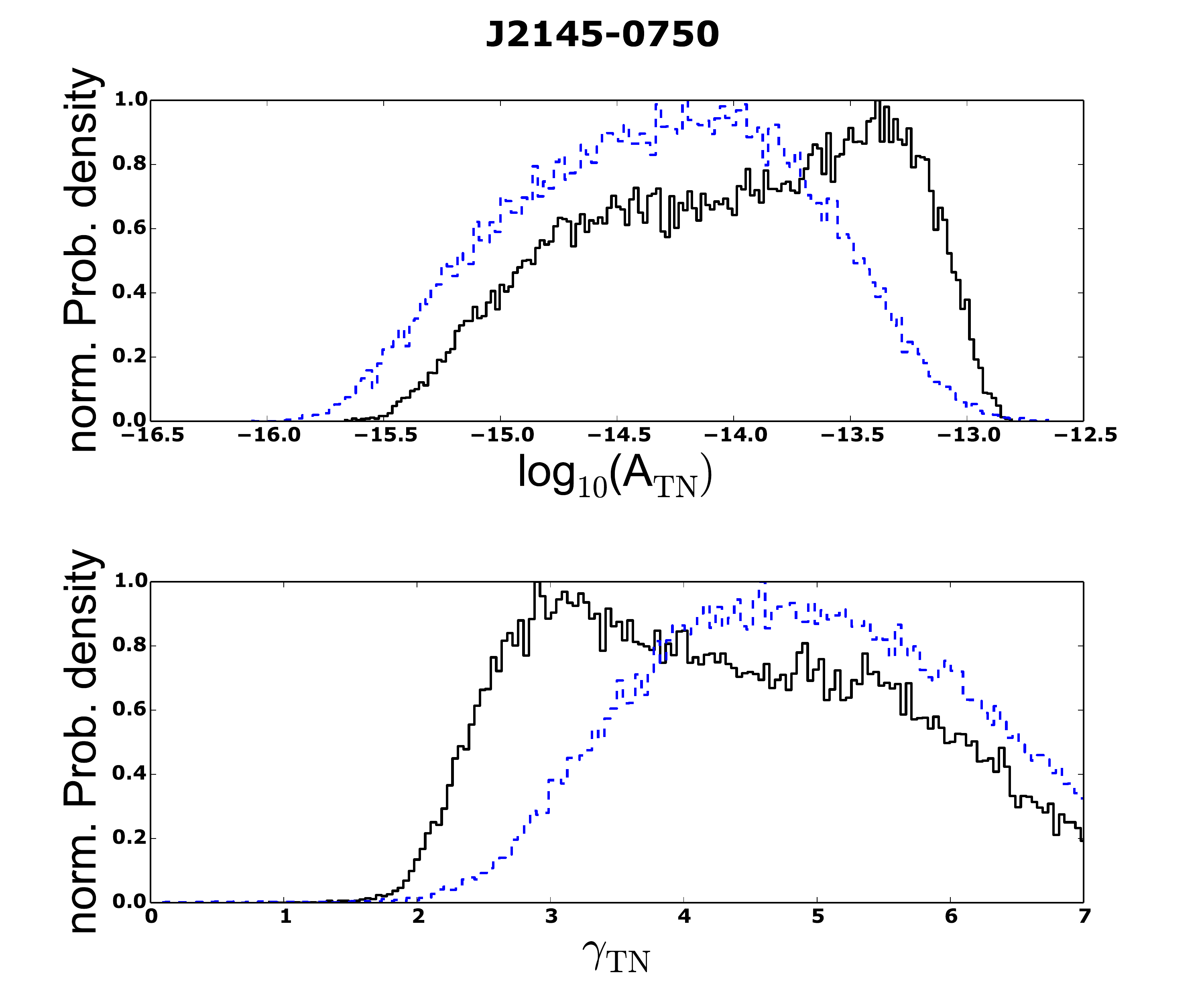}\\
\includegraphics[width=9cm]{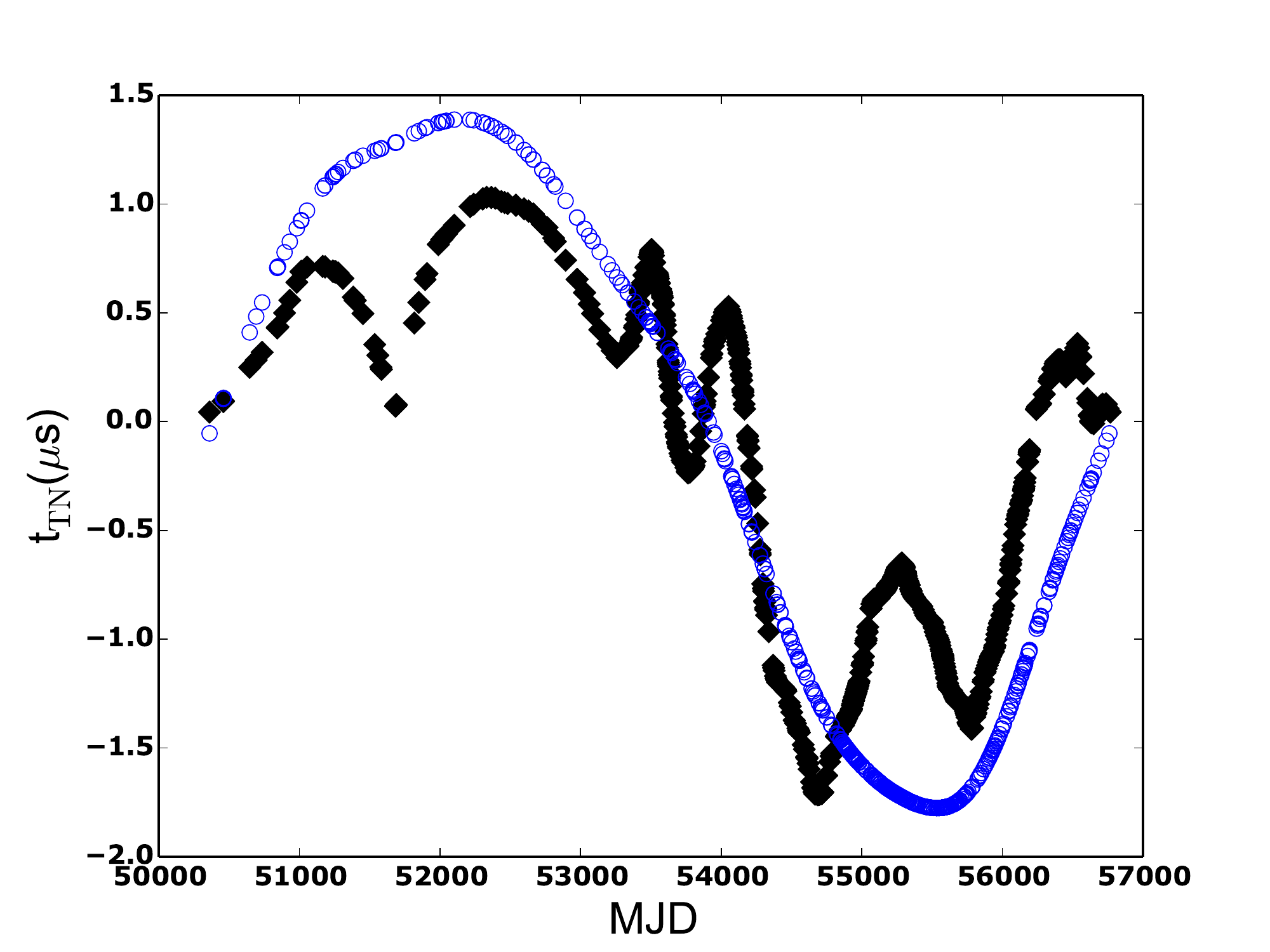} &
\includegraphics[width=9cm]{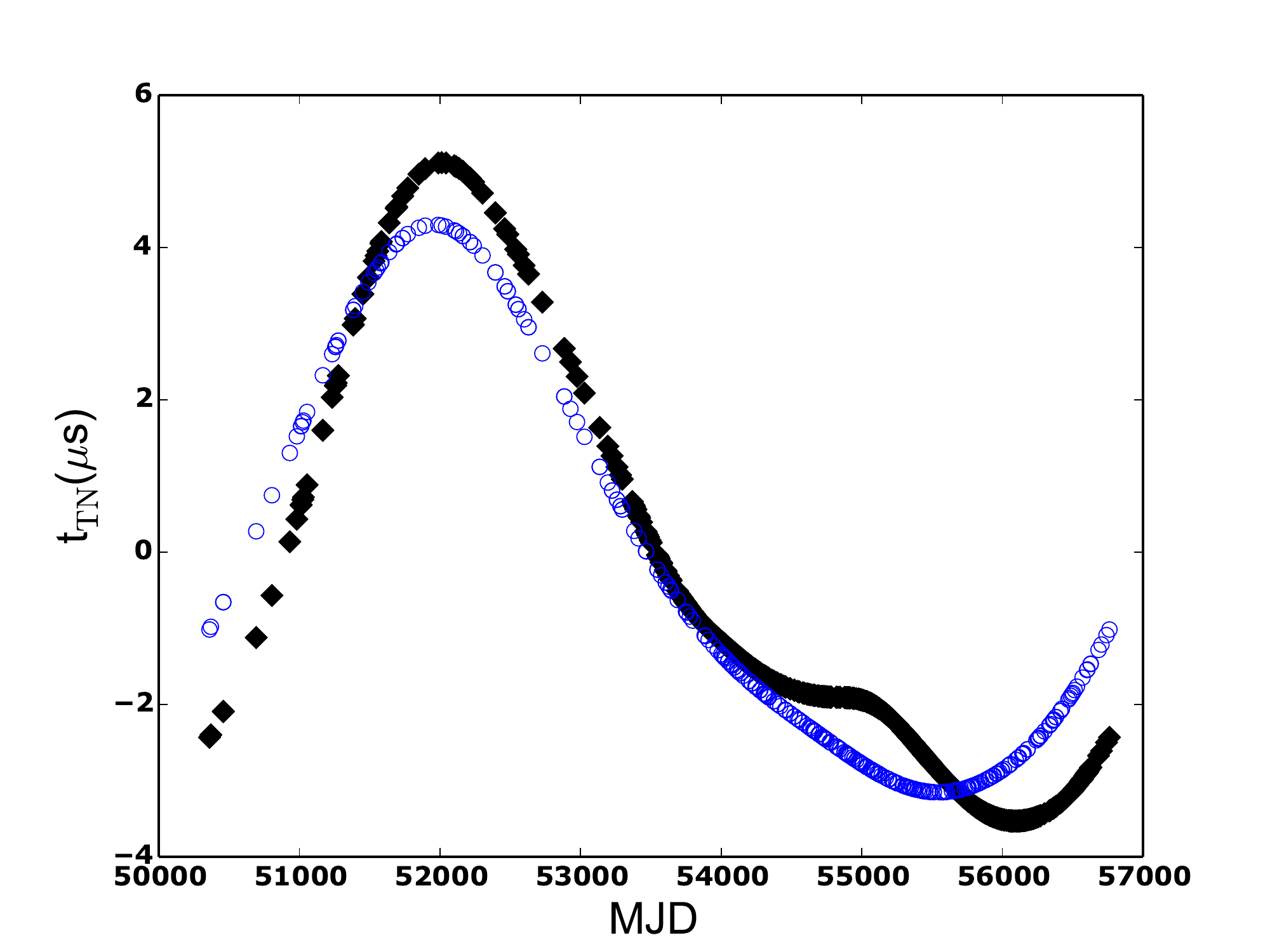}\\
\end{array}$
\caption{{\it Top panel:} Comparison of the 1-dimensional marginalised 
posterior probability 
distributions of the timing noise parameters when using 
the full EPTA data set (solid, black lines ) and the data subset which
does not include any NRT-BON data (blue, dashed lines). 
{\it Bottom panel:} Comparison of the timing noise waveforms (${\rm \bf t}_{\rm TN}$) 
when performing the noise analysis on the full EPTA data set (filled
black diamonds) and the data subset which excludes the NRT-BON data (open blue circles).
See Section~\ref{sect:systematics} for a discussion.}
\label{fig:InstrumentalPostPD}
\end{center} 
\end{figure*}

Our analysis shows evidence 
for TN specific to the NRT data. 
Figure~\ref{fig:InstrumentalPostPD} shows the PPDs for the TN parameters when using the full data set and 
when excluding the NRT data, and the respective
ML TN waveforms. 
For PSR J1022+1001 
the PPDs become significantly broader when excluding the NRT data. 
The mean value of the amplitude reduces by two 
orders of magnitude and the TN waveform becomes smoother, although the waveform has almost unchanged
peak-to-peak variations. The TN parameters PPDs of PSR J2145$-$0750 show a bimodality, which is not present 
when removing the NRT data. The two TN waveforms are almost identical, 
apart from the fact that the waveform of the full data set shows
a bump around MJD 56000, which is not present when removing the NRT data. 
These effects are most likely caused  either by additional noise in the NRT data from 
instrumental instabilities or by some additional non-instrumental noise component that 
only the NRT data are sensitive to, having indeed the highest precision TOAs. 
We stress that since we have assumed the TN to be stationary, 
the properties of instrumental noise during a specific time-interval 
can leak into the estimated TN waveform throughout the pulsar data set. We note
that there were known instrumental instabilities at the NRT 
during the period between MJD 54300-54500 (July 2007 to February 2008). 

This analysis can be better performed using the 
International Pulsar Timing Array (IPTA) data set (Verbiest et al., submitted) 
where data from another 3 telescopes 
are included, offering a larger amount of multi-telescope overlapping data. 
The presence of observing system-dependent noise is more extensively 
investigated in the paper examining the 
noise properties of the IPTA data set (Lentati et al., submitted.).  

\section{Timing Noise from Errors in Terrestrial Time Standards}
\label{sect:clockErr}
During pulsar timing observations, the TOAs are referenced against 
the local atomic clock (e.g. Hydrogen maser clock)
or a Global Positioning System (GPS) clock\footnote{This is the case for the NRT data}.
These clocks are stable on timescales of weeks, 
allowing good phase keeping 
(1-pulse-per-second signal) during observations. 
These clocks, however, show instabilities on timescales of months to years
and the TOAs recorded using them, are therefore not suitable for 
high-precision pulsar timing projects. This problem can be mitigated through the 
application of a series of corrections based on monitoring the offsets between pairs of clocks 
\citep[see e.g.][]{2006MNRAS.369..655H,2012MNRAS.427.2780H}. 

Cross-correlating the pulse profiles with the 
template profile references all arrival times to the same (arbitrary) phase, 
forming the topocentric TOAs. Unless the time-stamping was 
performed using a GPS clock, the TOAs are then converted 
to GPS-based Universal Coordinated Time (UTC) time, 
using clock correction files, created by calculating the difference between
the local atomic clock and the GPS times. This is then converted to UTC and subsequently 
to the International Atomic Time (TAI) standard. TAI is 
formed by the weighted average of the time-scales of several hundred atomic clocks around the world 
and subsequent frequency adjustments using primary frequency standards. 
These adjustments are made over timescales of years, 
a process known as ``steering''. As a result, TAI can have errors 
during the steering periods which are never 
retroactively corrected. For these reasons, 
for pulsar timing we use the corrections on TAI provided by the Bureau International 
des Poids et Mesures (BIPM)\footnote{http://www.bipm.org/}. 
These corrections are made through measuring offsets 
between various clock pairs to achieve the best possible precision and are regularly updated. 

Any possible remaining errors in the BIMP terrestrial time standard or 
error propagated to the TOAs by systematics when referencing the TOAs 
to the various time standards, will lead to a ``clock error'' signal, 
a monopolar correlated signal in the PTA 
sources, i.e. a signal with the same waveform in all pulsars and observing systems.
As discussed in \cite{2016MNRAS.455.4339T}, the mitigation of the clock error signal 
is of central importance in PTA efforts for GW detection. 
In this section, we search for a 
terrestrial clock error in the data set to determine
how much of the measured noise can be attributed to clock error noise.
Previously, \cite{2012MNRAS.427.2780H} presented their measurement of the clock 
error using data from the PPTA and discussed how pulsars can serve as an independent, 
non-terrestrial time standard.

\subsection{Methodology and results}
\label{subsec:clockMethod}
We use a maximum likelihood estimator to infer the clock error signal. 
The clock-error noise is modelled as a red-noise 
process power-law with power spectral density described by 
Eq.~\eqref{eq:RedNoiseFunc}, with amplitude 
$A_{\textrm{clk}}$ and spectral index $\gamma_{\textrm{clk}}$.  
Using the results on these parameters, 
we subsequently construct the ML signal waveform.
 
For this analysis, we set the TN parameters of the MSPs 
to the ML values from the Bayesian analysis with uniform pPDs on the TN amplitude
(as described in Section 3.2). In this way, we derive the ML solution for the 
clock-error noise with the higher possible amplitude, given our TN results.
We use the residuals 
after subtracting the ML DM variations signal as in Section 3.3, to focus on 
the TN only. 
The likelihood function is similar to Eq.~\eqref{eq:BayesLikFunc} but with the extension to multiple 
pulsars to investigate the clock signal, which is identical among all pulsars, as: 
\begin{equation}
	L\propto \frac{1}{\sqrt{|\rm \bf C|}} e^{-\frac{1}{2} \sum_{i,j,I,J} ({ 
	t}_{I,i}-{\tau}_{I,i}) {C}_{I,J, i, j}^{-1} ({ t}_{J,j}-{ \tau}_{J,j})}\,,
	\label{eq:liklimultipsr}
\end{equation}
where the index $I,J$ are for pulsars, and index $i,j$ are for the time epoch.
The total covariance matrix now includes the 
covariance matrix of the clock error signal, ${\bf C_{\textrm{clk}}}$, 
while not including the matrix of the DM 
variations such that, ${\bf C} = {\bf C_{\textrm{w}}} + {\bf C_{\textrm{TN}}} + {\bf C_{\textrm{clk}}}$. 
The intrinsic noise of pulsars is 
not correlated between pulsar pairs, so C$_{\textrm{w}\,I,J}=0$ and
C$_{\textrm{TN}\, I,J}=0$ for $I\neq J$. The clock error waveform is identical in all pulsars, 
therefore its covariance matrix components can be expressed as 
$C_{\rm clk\,I,J,i,j}=C_{\rm clk}(t_i-t_j)C_{\textrm{clk}\, I,J}$, 
with C$_{\textrm{clk}\, I,J}=1$ for all $I$,$J$ pairs. 
The likelihood function shows that for the estimation of the clock noise parameters we consider both the 
the clock error signal on the residuals of each pulsar (autocorrelation effect) 
and the cross-correlation of the residuals between pulsar pairs.

We make the linear approximation of the timing 
model as described in \cite{2009MNRAS.395.1005V}, i.e. 
considering linear deviations of the true timing parameter values, 
$\boldsymbol{\epsilon}$, from the 
least-square-fit timing model values, $\boldsymbol{\epsilon_{0}}$, via the linear relation 
$\delta(\boldsymbol{\epsilon})$ = $\boldsymbol{\epsilon}$ -  $\boldsymbol{\epsilon_{0}}$. 
We therefore substitute the expression for the residuals in 
Eq.~\eqref{eq:liklimultipsr}, ${\bf t} - \boldsymbol{\tau(\epsilon)}$, with $\delta{\bf t}=\delta{\bf t}_{post} - {\bf M}\delta\boldsymbol(\epsilon)$; $\delta {\bf t}_{post}$ 
are the post-fit timing residuals and ${\bf M}$ is the design matrix of the timing parameters.
We marginalise analytically over all timing parameters and get the reduced likelihood function:
\begin{equation}
\label{eq:reducedLikFunc1}
L\propto \frac{1}{\sqrt{|\rm \bf C|}} e^{-\frac{1}{2} \sum_{i,j,I,J} ({ 
	\delta t}_{I,i}) {C'}_{I,J, i, j}^{-1} ({ \delta t}_{J,j})}\,,
\end{equation}
with ${\bf C'}={\bf C}^{-1}-{\bf C}^{-1}{\bf M}({\bf M}^{\textrm{{\bf T}}}{\bf C}^{-1}{\bf M})^{-1}{\bf M}^{\textrm{{\bf T}}}{\bf C}^{-1}$. Going one
step further, we split the deterministic signal between 
that of parameters for which we want to marginalise over 
(usually the timing model parameters), $\delta{\bf t'}$
and the signal of parameters we assume unknowns of the likelihood 
function (see e.g. Section~\ref{subsec:SGW_limits}). 
We note the latter parameters with the vector lambda, and assume their waveforms to be 
described by the $\boldsymbol{S(\lambda)}$.
\begin{figure}
\begin{center}
\includegraphics[width=9.2cm]{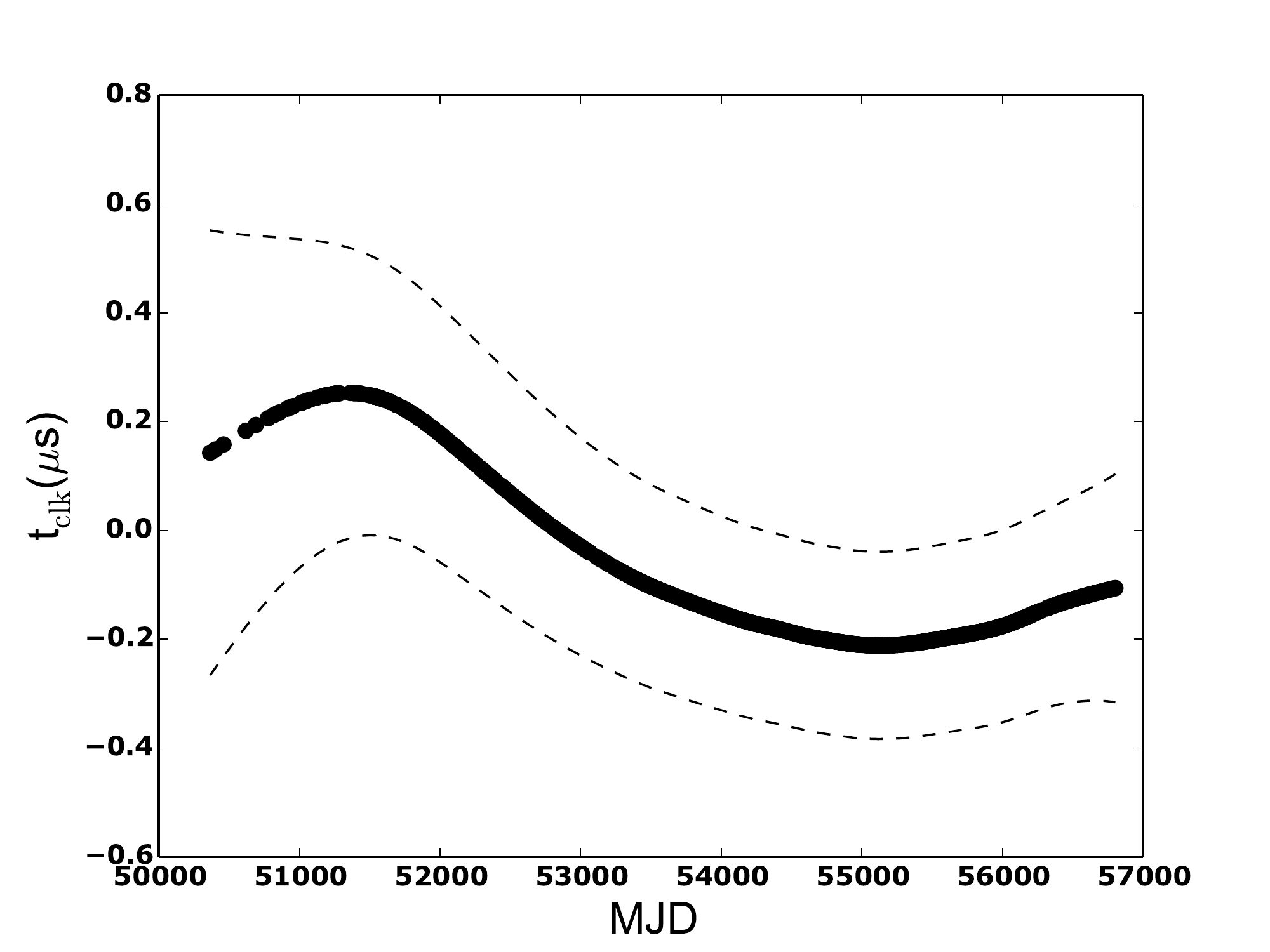}
\caption{The estimated waveform of the clock-error noise. 
The filled circles is the maximum likelihood waveform (${\rm \bf t}_{\rm clk}$). 
The dashed lines indicate the 68\% confidence intervals. 
For the estimation of the waveform, we used the upper limits 
for the values of the individual pulsar timing noise 
parameters providing upper limits for the clock error noise parameters.}
\label{fig:clockErr}
\end{center}  
\end{figure}
The likelihood function is then re-written as:
\begin{equation}
\label{eq:reducedLikFunc2}
L\propto \frac{1}{\sqrt{|\rm \bf C|}} e^{-\frac{1}{2} \sum_{i,j,I,J} ({ 
	\delta t'}_{I,i}-S(\lambda)_{I,i}) {C'}_{I,J, i, j}^{-1} ({ \delta t'}_{J,j}-S(\lambda)_{J,j})}\,,
\end{equation}
We sample A$_{clk}$ and $\gamma_{clk}$ over a uniform grid of values and search for the 
model that maximises the likelihood. The amplitude is sampled 
with values of log(A$_{clk}$) ranging from $-17.0$ to $-14.0$ with a step of 0.1, 
while the spectral index values range from 0.5 to 5 with a step of 0.1. Due to the 
large condition number of the clock error's covariance matrix, the individual 
likelihood computations are unstable. As such, the direct search for the ML 
solution with uniform grids produces non-desirable artefact (non-physical likelihood 
maxima). To avoid these effects, we performed a large number of
trials by dithering noise parameters with randomised offset values within each search grid. 
The likelihood value of the grid is taken to be the maximum of all trials.

To reduce the computational cost of the analysis we use the ``restricted data set'' proposed 
in \cite{2016MNRAS.455.1665B}. This consists of six MSPs from the full data set, which 
give 90\,\% of the sensitivity to CGWs. This ``restricted data set'' has also been used in the 
derivation of upper limits to the amplitude of GWs with the EPTA Legacy data set 
 \citep{2015MNRAS.453.2576L, 2015PhRvL.115d1101T, 2016MNRAS.455.1665B}.
The ``restricted data set'' contains the pulsars 
PSRs J0613$-$0200, J1012$+$5307, J1600$-$3053, J1713$+$0747, 
J1744$-$1134,  and J1909$-$3744. 

We find a ML solution at A$_{clk} =-15.2$ and 
$\gamma_{clk} =4.8$. We use these values to calculate the ML waveform of the 
signal, shown in Fig.~\ref{fig:clockErr}. The clock signal waveform can be 
estimated in the same way we did for TN:
\begin{equation}
	{\rm \bf t}_{\rm clk}={\rm \bf C}_{\rm clk} {\rm \bf C}^{-1} {\rm \bf t}\,.
	\label{eq:waveestimator}
\end{equation}
with uncertainties estimated as:
\begin{equation}
	{\rm \bf \sigma}_{\rm clk}={\rm \bf C}_{\rm clk} - {\rm \bf C}_{\rm clk} {\rm 
	\bf C}^{-1} {\rm \bf C}_{\rm clk}\label{eq:wavesigma}
\end{equation}
The upper limit clock error 
waveform has an rms value of 0.17$\mu s$. By integrating 
Eq.~\eqref{eq:RedNoiseFunc} from the lowest to the highest spectral frequency 
for the clock error noise we derive the average power of the signal. 
We can compare this to the average power of the noise for each MSP, which 
is calculated by adding the TN average power and the white noise 
average power ($S_{\textrm W}$, as in Eq.~\ref{eq:SA_RedNoise}).
We find that the contribution of the clock error noise to the total noise levels 
of the individual pulsars is less than 1\%.

\section{Effects of Timing Noise on prospects for GW detection}
\label{sect:GWprospect}

Various studies have examined the sensitivity of PTAs to GWB signals 
\citep[e.g.][]{2005ApJ...625L.123J,2012MNRAS.423.2642L,
2013CQGra..30v4015S}.
These studies focus on making detection significance estimations and 
projections based on analytic formulae or scaling laws, which are derived 
assuming a given detection technique. These 
estimates are usually made 
based on specific assumptions, such as: the TOAs are regularly sampled and
simultaneous across pulsars, that the measurement precision is constant and identical for all pulsars 
and the absence of 
low frequency noise. The detection significance is usually expressed as the precision 
by which the dimensionless amplitude can be measured for a given spectral index.

In this paper we make use of the Cr\'amer-Rao lower bound (CRLB) to investigate
the limitations of using the present data set in detecting GWs, both
for stochastic isotropic GWBs and CGWs from SMBHBs. 
The advantage of this method is that it takes into account
all the observational properties of the data, such as cadence, white
and TN levels, while still using analytic calculations that 
demand very few computational resources and does not require data 
simulations. The impact of the TN present in the data on the 
PTA's sensitivity to GWs can then be estimated by comparing the CRLB 
when using the full covariance matrix and when omitting the TN component. 

The CRLB states that, for any unbiased estimator, the variance is equal to or 
higher than the inverse of the Fisher information matrix, 
$\boldsymbol{\mathcal{I}}$. When the equality is valid the estimator is also 
``fully efficient'' \citep{ptms63}. As discussed in \cite{2008PhRvD..77d2001V}, the ML estimator 
(which we use in this analysis for the GW amplitude as described below) 
can achieve the bound in the high S/N regime. 
For the amplitude of GW signals, the CRLB represents the 
lowest uncertainties (in case of detection) 
or upper limits (in case on non-detection) any unbiased estimator can achieve. 
We note, that although the CRLB is underperformed by all unbiased estimator, in principle it can be 
outperformed by a biased estimator \citep{2008PhRvD..77d2001V}.
The interpretation of the bound as the amplitudes upper limit in the non-detection case warrants 
more caution, since by default it assumes we are outside the high S/N regime. Nevertheless, it is unlikely 
that other estimators can provide lower upper limits than the CRLB under the same assumptions. 
For the purpose of evaluating the role of TN on the data's sensitivity to GWs, we are primarily interested 
in the ratio of the CRLB when assuming only white noise in the data and when the TN is taken into account. 
Therefore, even if the individual CRLB results are not optimal, their ratio should be representative of the 
effects of TN. The CRLB calculated in the presence of TN are in fact comparable\footnote{Note that the CRLB
refers to the equivalent of a 68\% C.L. Typically, the 95\% C.L. is reported in the PTA literature for the 
amplitude of GWs.} to the 
amplitude limits derived in  \cite{2015MNRAS.453.2576L} and \cite{2016MNRAS.455.1665B} using 
more rigorous algorithms.

In its general form, the CRLB is formulated as follows. 
Given a likelihood function, $f(\boldsymbol{\lambda}, {\bf x})$, where ${\bf x}$ is the 
data and $\boldsymbol \lambda$ are the model parameters, the CRLB is:

\begin{equation}
\label{eq:CRLB1}
\textrm{Cov}(\boldsymbol{\lambda})=\langle 
\sigma_{\lambda_i}\sigma_{\lambda_j}\rangle \geq 
\boldsymbol{\mathcal{I}}_{ij}^{-1} ,
\end{equation}
where the indices i and j denote the different parameters and
$\mathcal{I}_{ij}$ is:
\begin{center}
\begin{equation}
\label{eq:CRLB2}
\mathcal{I}_{ij}=\left\langle\frac{\partial \ln 
f(\boldsymbol{x},\boldsymbol{\lambda})}{\partial
\lambda_i}\frac{\partial \ln f(\boldsymbol{x},\boldsymbol{\lambda})}{\partial \lambda_j}
\right\rangle \equiv-\left\langle\frac{\partial^2
\ln f(\boldsymbol{x},\boldsymbol{\lambda})}{\partial
\lambda_i\partial
\lambda_j}\right\rangle
\end{equation} 
\end{center}
It is well-known that $\mathcal{I}$ can be analytically calculated 
for Gaussian likelihood functions (as is  Eq.~\ref{eq:BayesLikFunc}), and 
results in the so-called Slepian-Bangs formula \citep{1954IRE...368S,1971PhDT..B}:
\begin{center}
\begin{equation}
\label{eq:CRLB3}
\mathcal{I}_{ij}=\frac{1}{2}\Big\{ {\rm  {\bf tr}}\Big\lbrack {\bf 
C}^{-1}\frac{\partial {\bf C}}{\partial \beta_i} {\bf C}^{-1}\frac{\partial 
{\bf C}}{\partial \beta_j}  \Big\rbrack + \frac{\partial {\bf S(\lambda)}}{\partial 
\lambda_i}{\bf C}^{-1} \frac{\partial {\bf S(\lambda)}}{\partial \lambda_j} 
\Big\}\,.
\end{equation}
\end{center} 
Here, $\beta_i$ are the model parameters describing the covariance matrix, $\lambda_i$, are the 
parameters describing the unknown waveform ${\bf S}$ and 
${\bf tr}$ is the matrix trace.

We make use of the same maximum likelihood estimator as in 
Section~\ref{sect:clockErr} (Eq.~\ref{eq:reducedLikFunc2}), 
but we replace the stochastic clock error 
signal with that of a stochastic and isotropic GWB and we set 
$\boldsymbol{S(\lambda)}$ to be the CGW signal from a single SMBHB, 
as detailed in Section~\ref{subsec:SGW_limits}. 
The likelihood function (Eq.~\ref{eq:reducedLikFunc2}) 
uses a total covariance matrix which 
includes the covariance matrix of the GWB, such that 
${\bf C} = {\bf C_{\textrm{w}}} + {\bf C_{\textrm{TN}}} + {\bf C_{gwb}}$. 
The GWB's covariance matrix, is dictated by the expected correlation 
coefficient in the residuals of every pulsar pair, described by the 
overlap reduction function \citep{2009PhRvD..79f2003F}, $\Gamma(\zeta)$, defined as: 
\begin{center}
\begin{equation}
\label{eq:HDcurve}
\Gamma(\zeta)=\frac{3}{8}\Big[ 1+\frac{\textrm{cos}\zeta_{IJ}}{3}+4(1-\textrm{cos} \zeta_{IJ})\textrm{ln}\Big(\textrm{sin}\frac{\zeta_{IJ}}{2} \Big) \Big] (1+\delta_{IJ}) . 
\end{equation}
\end{center}
Here, $\zeta_{IJ}$ is the angular separation between the $I$-th and the $J$-th pulsar, and $\delta_{IJ}$ is 
the Kronecker delta. In principle, both an Earth and a pulsar term contribute to the correlation and 
$\delta_{IJ}$ accounts for the latter. In the short-wavelength approximation, i.e. when the pulsars are 
separated from the Earth and from each other by many GW wavelengths, the 
overlap reduction function is also known as the Hellings-Downs curve \citep{1983ApJ...265L..39H}.
The components of the covariance matrix of the GWB are then expressed as  
${\rm  C}_{ {\rm gwb} I,J,i,j }={\rm  C}_{ {\rm gwb} }(t_{i}-t_{j}) \Gamma(\zeta_{IJ})$. 
As in the case of the clock error covariance matrix (Section~\ref{subsec:clockMethod}), 
the form of the covariance matrix
allows the calculation of the CRLB to include both the autocorrelation and 
cross-correlation effects of the GW.
 
For this analysis, we use the same six MSPs that we used to estimate the clock
error noise parameters in Section~\ref{subsec:clockMethod} and 
we set the TN properties to their 
ML values as estimated with the Bayesian pulsar noise 
analysis described in Section~\ref{sect:Bayes} and presented in Table~\ref{tab:Bayes}. 
As discussed in Section~\ref{subsec:clockMethod}, the estimation that the sensitivity 
loss to GWs when using this data subset is below $10\,\%$ was made for the case of CGWs. 
For low-frequency stochastic signals such as the GWB or the clock error signal, the sensitivity loss should be less. 
For CGWs, adding a pulsar with precise data only in part of its data span can increase the 
S/N of a detection significantly if the SMBHB orbit is fully sampled. 
In the case of the GWB, however, 
the targeted correlated signal must be found in cross-correlations of TOAs across a long time-span of order 
equal to the inverse of the GW frequency, 
with sufficient precision. We have verified this by calculating the CRLB for the GWB using 
40 MSPs and noting an improvement in the amplitude limit of order $2\,\%$. 
The scaling of the sensitivity to GWs with the number of MSPs, the S/N 
regime of the targeted signal and other factors have been studied elsewhere
\citep[e.g.][]{2012PhRvD..85d4034B, 2013CQGra..30v4015S} and is outside the 
scope of this work.

In order to focus on the impact of TN only, we mitigate the DM variations beforehand by 
subtracting the ML DM variations waveforms from the residuals.
For detailed derivations and astrophysical interpretations on GW limits using the EPTA 
Legacy data set, we refer the 
reader to \cite{2015MNRAS.453.2576L}, \cite{2015PhRvL.115d1101T} and \cite{2016MNRAS.455.1665B} 
for the cases of a stochastic and isotropic GWB, the anisotropy in the GWB and 
the CGW from individual SMBHBs respectively.

\subsection{Stochastic Gravitational-Wave Background}
\label{subsec:GWB_limits}
When estimating the CRLB for the GWB amplitude, the terms with partial derivatives of {\bf S} are zero 
and  Eq.~\eqref{eq:CRLB3} reduces to 

\begin{equation}
{\cal I}_{ij}=\frac{1}{2}{\rm  {\bf tr}}\Big\lbrack {\bf 
C}^{-1}\frac{\partial {\bf C}}{\partial \beta_i} {\bf C}^{-1}\frac{\partial 
{\bf C}}{\partial \beta_j}  \Big\rbrack .
\end{equation}
We calculate the CRLB for the GWB amplitude, keeping each time 
the GWB spectral index fixed. We do so for a range of spectral indices, from $-2$ to 1, 
which covers GWB signals often 
discussed in PTAs literature, e.g. 
from SMBHBs, cosmic strings and the relic GWB from the inflationary era.

This simplified 
approach intends to provide an understanding of the difficulties the TN imposes on the detection 
of the various GWBs probed by PTAs. It is not exhaustive, since each of these 
GWBs can in general have a range of possible spectral index values. 
In the case of SMBHBs, 
this depends on the orbital eccentricities and whether the 
SMBHBs are coupled to their stellar and gaseous environment or they are driven 
by GW emission only \citep{2013CQGra..30v4014S}. The often used power-law 
index of $-2/3$ refers to 
circular, GW-driven SMBHBs \citep{1995ApJ...446..543R,2003ApJ...583..616J}. 
Strong environment coupling and 
high orbital eccentricities can cause a turnover of the 
spectrum at low-frequencies \citep[e.g. Fig.~2 in][]{2013CQGra..30v4014S}.
The value -7/6 we have used for the spectral index of the cosmic string 
GWB has been analytically derived using 
a simplified approximation of the loop number density and 
assuming cusp emission \citep[e.g.][]{2005PhRvD..71f3510D}. 
However, especially in the frequencies probed by PTAs, 
a wide range of spectral indices is possible, depending 
on some characteristic parameters used to describe the 
evolution of the network and the details of the dominant 
GW emission mechanism, 
and one  typically sets limits on the amplitude for a 
range of these parameters \citep{2012PhRvD..85l2003S}
For the cosmological relic GWB, a spectral index of $-1$ is often cited
\citep{2005PhyU...48.1235G}. For more details on the sources of the various GWBs and 
details on the derivation of amplitude limits as function of the spectral index and 
other physical parameters, we refer the reader to \cite{2015MNRAS.453.2576L,2015arXiv150803024A} 

\begin{figure*}
\begin{center}$
\begin{array}{c c}
\includegraphics[width=9.0cm, angle=0]{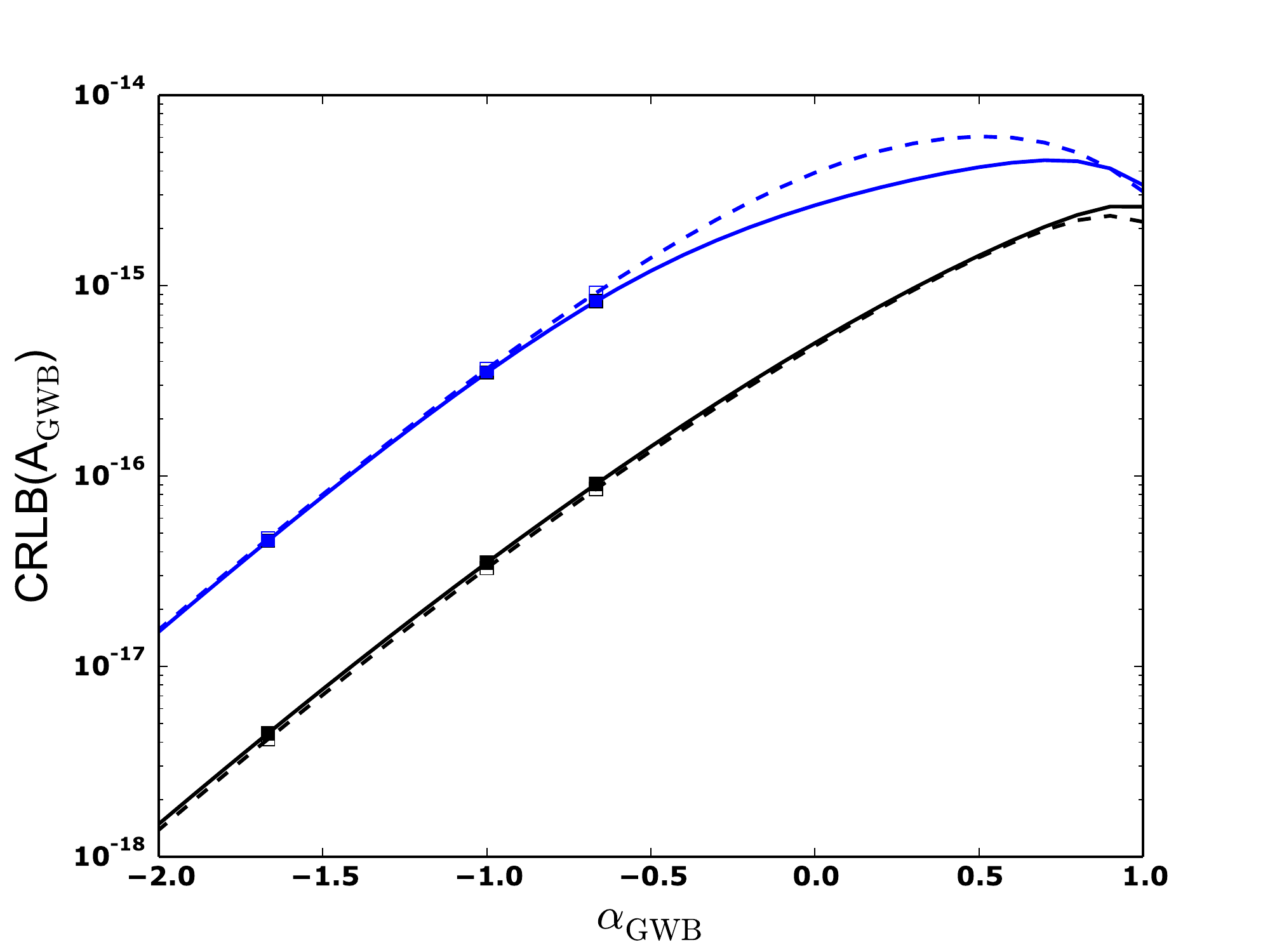} &
\includegraphics[width=9.0cm, angle=0]{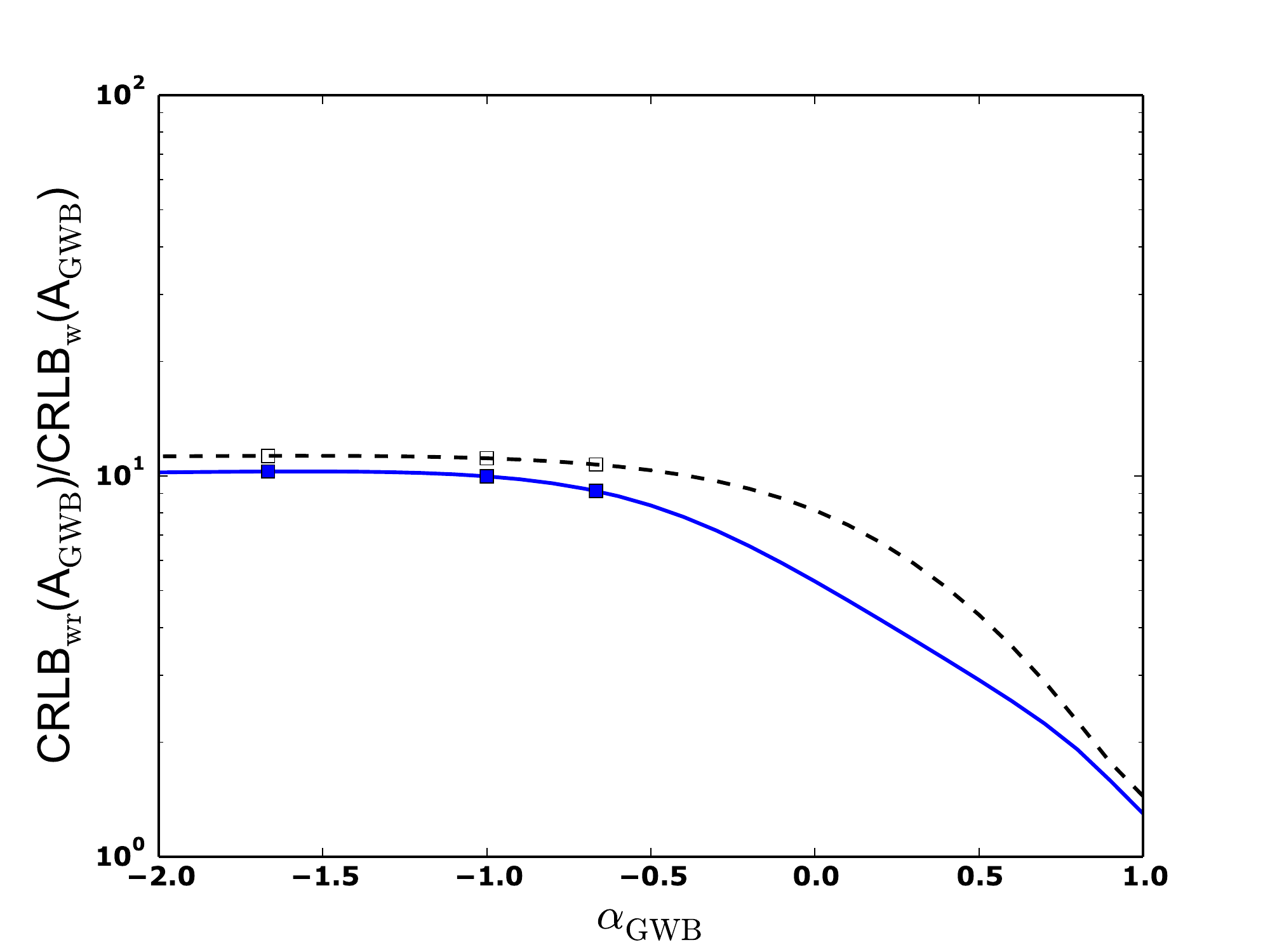} \\
\end{array}$
\caption{Cr\'{a}mer-Rao lower bounds (CRLBs) for the strain amplitude of GWBs, $A_{GWB}$ for a range of spectral indices, 
$\alpha_{GWB}$. Squares denote the values shown in Table~\ref{tab:CRLBGWB}. 
{\it Left panel:} CRLBs calculated bounds using the TN maximum likelihood parameters from the
Bayesian analysis using log-uniform priors on the TN amplitude (solid lines and filled squares) 
and using uniform priors on the TN amplitude (upper limits on 
TN; dashed lined with open squares). Blue symbols are for limits calculated assuming  timing and white noise, while black symbols when only the white noise levels are taken into account.  
{\it Right panel:} The ratio of the CRLBs for when assuming timing and white-noise and white noise only in
the data. The blue solid line is when using the timing noise properties the analysis with log-uniform priors 
and the while the black dashed line is for the analysis with uniform priors.} 
\label{fig:CRLB_GWB}
\end{center} 
\end{figure*}

The CRLBs are calculated using the TN parameters from the two Bayesian analyses, 
using different types of pPDs on the TN noise amplitude. 
For each set of TN results, we calculate the CRLB for two cases, namely 
assuming the presence of the measured white and TN, or assuming
only the measured white noise levels, and finally, calculate their ratios.  
Figure~\ref{fig:CRLB_GWB} shows the results for both cases.
The results for the spectral indices representative of GWBs 
from SMBHBs, cosmic strings and relic 
GWs are presented in Table~\ref{tab:CRLBGWB}. 
The improvement factor on the lower bound when assuming no 
TN in the data is always more than an order of magnitude, ranging from 9.1 to 11.4.
These results demonstrate how strongly TN can reduce the data's sensitivity to GWs. To stress this even 
further, we note that the upper limits on the GWB amplitude by SMBHBs (spectral 
index $-2/3$) by PTAs have improved by a factor of ten over the past ten years. 

\begin{table}
\centering
\caption{Results for the Cr\'{a}mer-Rao bound (CRLB) on a 
GWB for the expected signals from SMBHBs ($\alpha=-2/3$), cosmic
strings ($\alpha=-7/6$) and cosmological relic GWs ($\alpha=-1$) (see Section~\ref{subsec:GWB_limits}
for a discussion on the noted spectral indices). We tabulate the CRLB assuming both the 
measured white- and timing-noise levels, ($A_{GWB_{\textrm{wr}}}$) for measured and upper limit values
and white-noise levels only ($A_{GWB_{\textrm{w}}}$).}
\begin{tabular}{c c c c c}
\hline
 & Max. likelihood TN \\
\hline
$\alpha_{GWB}$ & $A_{GWB_{\textrm{wr}}}$ & $A_{GWB_{\textrm{w}}}$ & $\frac{A_{GWB_{\textrm{wr}}}}
{A_{GWB_{\textrm{w}}}}$\\
\hline
-2/3 & 8.3$\times10^{-16}$ & 9.1$\times10^{-17}$ & 9.1\\
-7/6 & 4.6$\times10^{-17}$ & 4.4$\times10^{-18}$ & 10.3\\
-1 & 3.5$\times10^{-16}$ & $3.5\times10^{-18}$& 10.0\\
\hline
 & Red noise upper limits\\
\hline
-2/3 & 9.2$\times10^{-16}$ & 8.5$\times10^{-17}$ & 10.7\\
-7/6 & 4.7$\times10^{-17}$ & 4.1$\times10^{-18}$ & 11.4\\
-1 & 3.7$\times10^{-16}$ & 3.3$\times10^{-17}$& 11.1\\
\hline
\end{tabular}
\label{tab:CRLBGWB}
\end{table}

\subsection{Gravitational Waves from single SMBHBs}
\label{subsec:SGW_limits}

\begin{figure*}
\begin{center}$
\begin{array}{c c}
\includegraphics[width=9.0cm, angle=0]{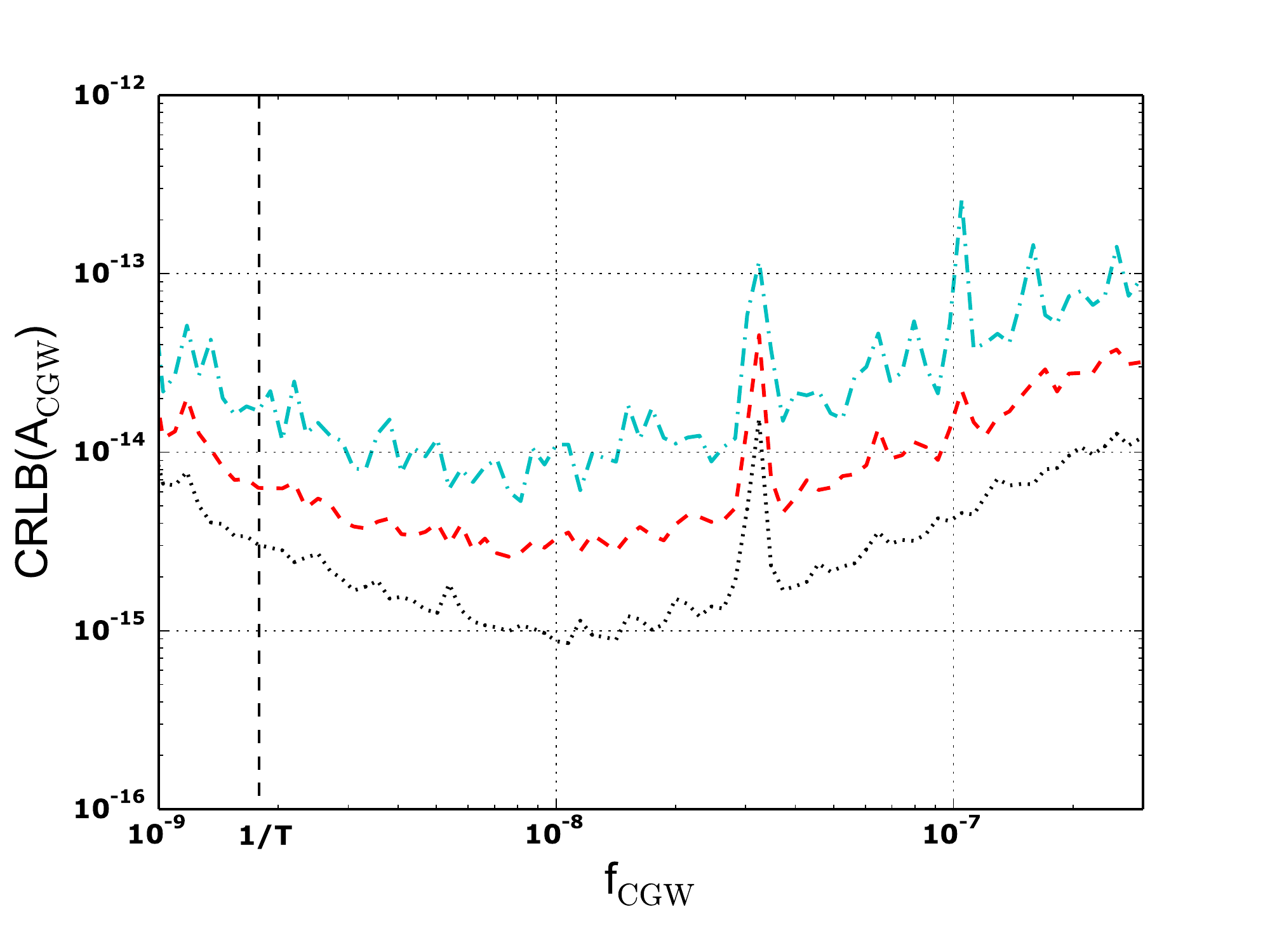} &
\includegraphics[width=9.0cm, angle=0]{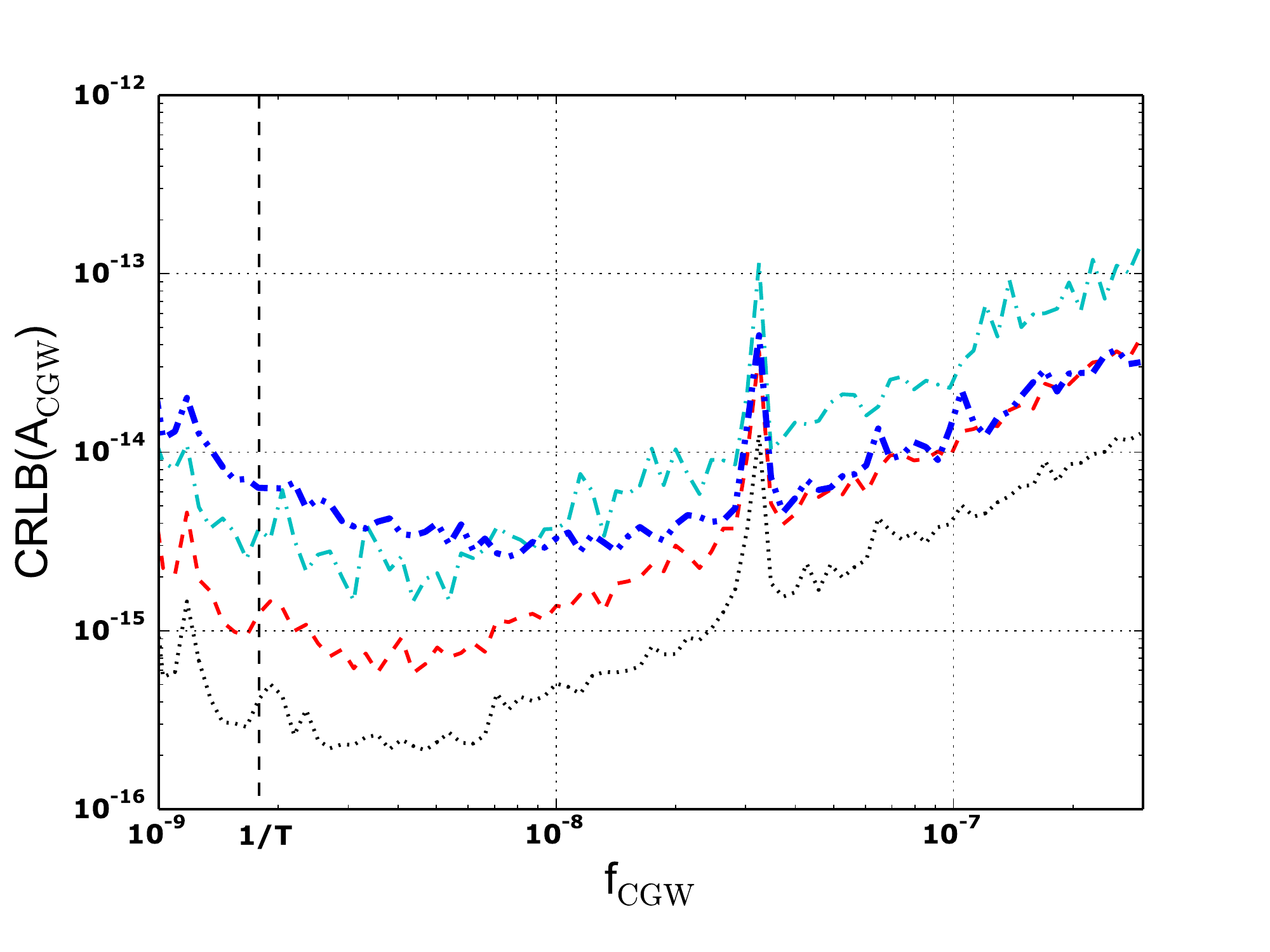} \\
\end{array}$
\caption{Results for the Cr\'{a}mer-Rao lower bound (CRLB) on the strain amplitude of 
continuous GWs, $A_{CGW}$, against the CGW frequency, $\textrm{f}_{\textrm{CGW}}$, 
from resolvable SMBHBs with circular orbits and without measurable frequency chirping. 
The different curves are for the cases 
where the SMBHB is at the sky location where the PTA has the maximum 
(cyan, dot-dashed lines) and minimum (black, dotted lines) 
sensitivity, and the average of all positions on the sky (red, solid lines). The vertical line show the 
frequency resolution of the PTA, 1/T, where T is time-span of the pulsar with the longest data set. 
{\it Left panel:} Sensitivity curves when accounting for the 
white and the timing noise of the data. {\it Right panel:} Sensitivity curves when only 
accounting the white noise of the data. The additional blue, thick double dot-dashed line is the 
case for mean PTA sensitivity when including the timing noise as in the left panel (red, solid line) for 
better comparison.} 
\label{fig:CRLB_CGW}
\end{center} 
\end{figure*}

Here we focus on CGWs from resolvable, GW-driven SMBHBs 
with circular orbits and without measurable frequency evolution of the signal over the observing interval
due to energy loss from the binary by GW emission (an effect known as frequency chirping, see e.g.  \cite{2009ARA&A..47..107H}). 
The waveform ($\rm \bf S$) of CGWs has been calculated 
by many independent studies \citep[e.g.][]{1987GReGr..19.1101W, 2006LRR.....9....4B,2009ARA&A..47..107H}. 
For each SMBHB, the 
waveform is characterised by seven parameters, namely
the GW amplitude, frequency and phase, the SMBHB's sky co-ordinates (right ascension and declination), orbital 
inclination, and direction of the binary's ascending node on the sky. Clearly, the 
terms with partial derivatives of {\bf C} are zero for the single SMBHB signal 
and Eq.~\eqref{eq:CRLB3} reduces to

\begin{equation}
{\cal I}_{ij}=\frac{1}{2}\frac{\partial {\bf S}}{\partial \lambda_i}{\bf 
C}^{-1} \frac{\partial {\bf S}}{\partial \lambda_j}\,.
	\label{eq:CRLB_single}
\end{equation}
Due to the seven parameters, the covariance matrix for the single GW source is a 
$7\times7$ matrix. The CRLB of the single source amplitude depends on the GW 
frequency, source position, orbital inclination and orientation. It has been 
shown \citep{2013CQGra..30v4016L} that the precision estimation of the GW 
source position using CRLB would be 
poor, due to the lack of a unique un-biased estimator for the single source 
problem. The statistics of the amplitude estimator, on the 
other hand, can be well described by the CRLB, which determines the sensitivity of 
a PTA as function of frequency. The sensitivity depends on the GW source 
position. We estimate the CRLB for three scenarios: placing the SMBHB at the 
sky position where the PTA has the minimum and maximum sensitivity as well
as the average of all positions on the sky. Our results are given in 
Fig.~\ref{fig:CRLB_CGW}.  
The low-frequency sensitivity extends to values lower than 
the frequency resolution (1/T) because the GW low frequency signal still leaks power into 
the observing window after the pulsars' spin and spin-down fitting. This 
causes the curve to rise below the frequency resolution. The rise of the curve at high GW frequencies 
is due to the PTA frequency response, as the GW induced timing residuals are the time 
integral of the GW strain. The peak at 1~yr$^{-1}$ (3.17$\times10^{-8}$~Hz) 
is caused by the pulsar sky position fitting. 

The improvement in the PTA sensitivity at low frequencies is obvious from Fig.~\ref{fig:CRLB_CGW}. 
One can clearly notice how the presence of TN flattens 
the sensitivity below $\sim$10~nHz, which, in contrast, keeps 
improving in the case of timing data free of TN. In the absence of TN, the sensitivity at 
low GW frequencies is only limited by the PTA's frequency resolution. 
Table~\ref{tab:CRLBCGW} summarises the CRLBs for the CGWs 
amplitude at frequencies of 5 and 7~nHz and the improvement factors to the sensitivity when the data do not 
have TN, which range from 2.3 to 5.6.

\begin{table}
\centering
\caption{Results for the Cr\'{a}mer-Rao lower bound (CRLB) on the strain amplitude of continuous GWs 
from resolvable SMBHBs with circular orbits and without measurable frequency chirping. 
We quote the limits for the cases when the SMBHB is at the sky location where the PTA has the maximum 
(max) and minimum (min) sensitivity, and the average of all sky positions (avg) at GW frequencies 
of 5 and 7~nHz. For each case we quote the limits when accounting for the white and the TN of the data, 
$A_{\textrm{CGW}_{\textrm{wr}}}$ and for the white noise only, $A_{\textrm{CGW}_{\textrm{w}}}$. 
The last column shows the ratio of the limits for these two cases.}
\begin{tabular}{c c c c c}
\hline
\hline
GW freq. & $A_{\textrm{CGW}_{\textrm{wr}}}$ & $A_{\textrm{CGW}_{\textrm{w}}}$ & $\frac{A_{\textrm{CGW}_{\textrm{wr}}}}{A_{\textrm{CGW}_{\textrm{w}}}}$\\
 (nHz) &  &   & \\
\hline
& Max PTA & sensitivity &\\
\hline
5 & 1.2$\times10^{-14}$ & 2.1$\times10^{-15}$ & 5.6\\
7 & 9.1$\times10^{-15}$ & 3.8$\times10^{-15}$ & 2.4\\
\hline
& Avg PTA & sensitivity &\\
\hline
5 & 4.0$\times10^{-15}$ & 8.1$\times10^{-16}$ & 5.0\\
7 & 2.7$\times10^{-15}$ & 1.1$\times10^{-15}$ & 2.4\\
\hline
& Min PTA & sensitivity &\\
\hline
5 & 1.3$\times10^{-15}$ & $2.4\times10^{-16}$& 5.3\\
7 & 1.0$\times10^{-15}$ & $4.4\times10^{-16}$& 2.3\\
\hline
\hline
\end{tabular}
\label{tab:CRLBCGW}
\end{table}

\section{Conclusions}
\label{sect:Conclusions}
In this paper, we have characterised the noise properties for 42 MSPs, using the EPTA 
Legacy data set. While the central focus is on the timing noise properties, we have also characterised 
the white noise properties of the data. 
The long time-spans of the pulsar data sets (the shortest being 6.9 years and 
the longest 24.1 years long) of high-quality 
timing data, are especially valuable for determining the timing noise. In order 
to increase our confidence in the results, we have employed two established methods, one based 
on Bayesian and the other one on power-spectral analysis. 
We used the Bayesian pulsar timing analysis package 
\tn{} to simultaneously determine the time-correlated timing noise, 
DM variations and uncorrelated noise (white-noise) properties. In order to focus the comparison 
between the methods on the timing noise characterisation, we used the 
maximum likelihood \tn{} results on DM variations and white-noise parameters as a priori known information 
when performing the frequentist analysis, based on a developed power-spectral analysis code described in this paper. For pulsars with statistically significant timing noise measurements, 
the two methods give statistically consistent results. 

The lack of sufficient multi-frequency data in 17 pulsars where timing noise is detected
leads to strong covariances between the timing noise and DM 
variations, causing the posterior distributions of the 
noise parameters derived from the Bayesian analysis to have probability tails 
extending to $\pm \infty$. These reflect the small probabilities of 
the noise amplitude to be zero, causing some deviations between the maximum likelihood and mean 
values of the parameters. The values of the ML and mean parameters as well as the parameter 
values estimated with the power-spectral analysis, are still however statistically consistent. 
Upper limit analysis is performed in these cases to set robust upper limits on the timing noise amplitude. 

Our analysis shows evidence of timing noise specific to the NRT data, which are likely linked to improper 
polarisation calibration in a roughly six-month-long epoch. 
We have also placed an upper limit on clock-error timing noise and 
find that it contributes at most 1\% to the total noise 
in the MSPs under examination. Finally, we assessed the role 
of timing noise in the efforts for GW detection using PTAs. 
We did so by estimating the Cr\'amer-Rao lower bound on the 
strain amplitude of a stochastic GWB and 
CGWs from resolvable SMBHBs, 
accounting only for the
measured white noise first and then adding the measured timing noise
properties. We find that, for GWBs, the timing noise in this data set 
reduces the sensitivity of this data set by a factor of 9.1 to 11.4, 
depending on the GWB spectral index. For CGWs, the 
sensitivity reduces by a factor of 2.3 to 5.6, depending on the GW 
frequency and the sky position of the SMBHB 
with respect to the sky position where the PTA is most sensitive. 

The results of this paper stress in a clear way the 
imperative need of PTAs to improve the noise characterisation and mitigation techniques and the 
development of good observing and data reduction practices to avoid introducing timing noise due to 
systematics. It also demonstrates the demand for new discoveries of MSPs that are not only bright, but also 
exhibit stable rotation. The rotational stability of pulsars can only be evaluated via timing-noise characterisation 
on data sets that are at least five years long, making the long-term follow-up timing observations 
of newly discovered MSPs essential for PTA observing campaigns. 

\section*{Acknowledgments}\label{}
We are grateful to Mike Keith, Bill Coles and George Hobbs for useful discussions. 

Part of this work is based on observations with the 100-m telescope of the Max-Planck-Institut f{\"u}r 
Radioastronomie (MPIfR) at Effelsberg. The Nan\c cay Radio Observatory is operated by the Paris Observatory, 
associated to the French Centre National de la Recherche Scientifique (CNRS). We acknowledge financial 
support from ``Programme National de Cosmologie and Galaxies'' (PNCG) 
of CNRS/INSU, France. Pulsar research at 
the Jodrell Bank Centre for Astrophysics and the observations using the Lovell Telescope is supported by a 
consolidated grant from the STFC in the UK. The Westerbork Synthesis Radio Telescope is operated by the 
Netherlands Institute for Radio Astronomy (ASTRON) with support from The Netherlands Foundation for Scientific 
Research NWO.

RNC acknowledges the support of the International Max Planck Research School Bonn/Cologne and the Bonn-
Cologne Graduate School. KJL gratefully acknowledge support from National Basic Research Program of
China, 973 Program, 2015CB857101 and NSFC 11373011. 
PL acknowledges the support of the International Max 
Planck Research School Bonn/Cologne. 
SO is supported by the Alexander von Humboldt Foundation. JWTH 
acknowledges funding from an NWO Vidi fellowship and from the European Research Council under the 
European Union's Seventh Framework Programme (FP/2007-2013) / ERC Starting Grant agreement nr. 337062 
(``DRAGNET''). CMFM was 
supported by a Marie Curie International Outgoing Fellowship within the 7th European Community Framework 
Programme. AS is supported by the Royal Society. 
This research was in part supported by ST's appointment to the 
NASA Postdoctoral Program at the 
Jet Propulsion Laboratory, administered by Oak Ridge Associated Universities through a contract with NASA. 
RvH is supported by NASA Einstein Fellowship grant PF3-140116. 

\bibliography{EPTA_DR1_Noise_paper_rev2.bib}            
\bibliographystyle{mnras}

\end{document}